\documentclass[12pt,preprint]{aastex}

\usepackage{lineno} 

\usepackage{paralist}
\usepackage{booktabs} 
\usepackage{multirow}
\usepackage{siunitx}
\usepackage{amsmath}

\newcommand{\e}{\mathrm{e}}

\shorttitle{Three Year GRB Cascade Paper}
\shortauthors{M.~G.~Aartsen et al.}

\begin{document}


\author{
IceCube Collaboration:
M.~G.~Aartsen\altaffilmark{1},
K.~Abraham\altaffilmark{2},
M.~Ackermann\altaffilmark{3},
J.~Adams\altaffilmark{4},
J.~A.~Aguilar\altaffilmark{5},
M.~Ahlers\altaffilmark{6},
M.~Ahrens\altaffilmark{7},
D.~Altmann\altaffilmark{8},
T.~Anderson\altaffilmark{9},
I.~Ansseau\altaffilmark{5}
% ...
}
\altaffiltext{1}{Department of Physics, University of Adelaide, Adelaide, 5005, Australia}
\altaffiltext{2}{Technische Universit\"at M\"unchen, D-85748 Garching, Germany}
\altaffiltext{3}{DESY, D-15735 Zeuthen, Germany}
\altaffiltext{4}{Dept.~of Physics and Astronomy, University of Canterbury, Private Bag 4800, Christchurch, New Zealand}
\altaffiltext{5}{Universit\'e Libre de Bruxelles, Science Faculty CP230, B-1050 Brussels, Belgium}
\altaffiltext{6}{Dept.~of Physics and Wisconsin IceCube Particle Astrophysics Center, University of Wisconsin, Madison, WI 53706, USA}
\altaffiltext{7}{Oskar Klein Centre and Dept.~of Physics, Stockholm University, SE-10691 Stockholm, Sweden}
\altaffiltext{8}{Erlangen Centre for Astroparticle Physics, Friedrich-Alexander-Universit\"at Erlangen-N\"urnberg, D-91058 Erlangen, Germany}
\altaffiltext{9}{Dept.~of Physics, Pennsylvania State University, University Park, PA 16802, USA}
% ...

\title{An All-Sky Search for Three Flavors of Neutrinos from Gamma-Ray Bursts with the IceCube Neutrino Observatory}





\begin{abstract}
We present the results and methodology of a search for neutrinos produced in the decay of charged pions created in interactions between protons and gamma-rays during the prompt emission of 807 gamma-ray bursts (GRBs) over the entire sky.  This three-year search is the first in IceCube for shower-like Cherenkov light patterns from electron, muon, and tau neutrinos correlated with GRBs.  We detect five low-significance events correlated with five GRBs.  These events are consistent with the background expectation from atmospheric muons and neutrinos.  The results of this search in combination with those of IceCube's four years of searches for track-like Cherenkov light patterns from muon neutrinos correlated with Northern-Hemisphere GRBs produce limits that tightly constrain current models of neutrino and ultra high energy cosmic ray production in GRB fireballs.
\end{abstract}


\keywords{gamma-ray bursts: general, methods: data analysis, neutrinos, telescopes}


\section{Introduction}

Ultra high energy cosmic rays (UHECRs), defined by energy greater than $10^{18}$ eV, have been observed for decades \citep{zatsepin1966,hires2008,auger2010}, but their sources remain unknown.  Gamma-ray bursts (GRBs) are among the most plausible candidates for these particles' origins.  If this hypothesis is true, neutrinos would also be produced in $p\gamma$ interactions at these sources.  While the charged cosmic rays are deflected by galactic and intergalactic magnetic fields, neutrinos travel unimpeded through the universe and disclose their source directions.  Detection of high energy neutrinos correlated with gamma-ray photons from a GRB would provide evidence of hadronic interaction in these powerful phenomena and confirm their role in UHECR production.

GRBs are the brightest electromagnetic explosions in the universe and are observed by dedicated spacecraft detectors at an average rate of about one per day \citep{meszaros2006,fermicat}.  GRB locations are distributed isotropically at cosmic distances.  Their prompt gamma-ray emissions exhibit diverse light curves and durations, lasting from milliseconds to hours.  The origins of these extremely energetic phenomena remain unknown, but their spectra have been studied extensively over a wide range of energies \citep{fermispec, fermilatcat2013, swiftbatcat2011}.  The \textit{Fermi} detector covers seven decades of energy in gamma-rays with its two instruments and has observed photons with source-frame-corrected energies above 10 GeV early in the prompt phase of some bursts \citep{fermiicrc2015}.  These measurements support efficient particle acceleration in GRBs.

The prevailing phenomenology that successfully describes GRB observations is that of a relativistically expanding fireball of electrons, photons, and protons \citep{piran,meszaros2006,fox2006}.  The initially opaque fireball plasma expands by radiation pressure until it becomes optically thin and emits the observed gamma-rays.  During this expansion, kinetic energy is dissipated through internal shock fronts that accelerate electrons and protons via the Fermi mechanism to the observed gamma-ray and UHECR energies \citep{waxman1995,vietri1997}.  These high energy protons will interact with gamma-rays radiated by electrons and create neutrinos, for example through the delta-resonance:

\begin{equation}
p + \gamma \rightarrow \Delta^{+} \rightarrow \pi^{+} + n \rightarrow e^{+} + \nu_{e} + \bar{\nu}_{\mu} + \nu_{\mu} + n
\label{deltares}
\end{equation}

To date, no neutrino signal has been detected in searches for muon neutrinos from GRBs in multiple years of data from AMANDA, the partially instrumented IceCube, and the completed IceCube detector \citep{amandacscd2007,ic22paper,grbic40paper,ic59nature,grb4year}, nor in four years of data by the ANTARES collaboration \citep{ANTARESGRBcscd,ANTARESGRB2013,ANTARESGRB2013b}.  High energy $\nu_{\mu}$ charged-current interactions produce high energy muons that manifest as extended Cherenkov light patterns in the South Pole glacial ice, referred to as ``tracks"; and Southern Hemisphere bursts were often excluded from searches for this signal in order to remove the dominant cosmic-ray-induced muon background.  Adding the low-background ``shower" channel from interactions other than charged-current $\nu_{\mu}$ gives enhanced sensitivity to such bursts, improving the sensitivity of the overall correlation analysis.  Both the shower and track analyses are sensitive to the combined flux of neutrinos and anti-neutrinos and are unable to distinguish between them.

This paper is ordered as follows.  In Section \ref{sec:promptgrbnuspec}, we describe the neutrino spectra predicted by the different fireball models on which we place limits.  In Section \ref{sec:detector}, we describe the IceCube detector and data acquisition system.  We discuss the simulation and reconstruction of events in IceCube in Section \ref{sec:simandreco}.  We detail the event selection techniques and likelihood analysis in Sections \ref{sec:eventselect} and \ref{sec:llh}.  Finally, in Section \ref{sec:results} we present the results of this all-sky three-flavor shower search in combination with those from the $\nu_{\mu}$ track searches, and conclude in Section \ref{sec:outlook}.

\section{Prompt GRB Neutrino Predictions}
\label{sec:promptgrbnuspec}

In this search for GRB neutrinos, we examine data during the time of gamma-ray emission reported by any satellite for each burst.  We do not consider possible precursor \citep{razzaque2003} or afterglow \citep{wb2000,murase2006,burrows2007} neutrino emission far outside of this prompt window.  In Section \ref{sec:results} we place limits on two classes of GRB prompt neutrino flux predictions: models normalized to the observed UHECR flux \citep{katz2009energy} and models normalized to the observed gamma-ray flux for each burst.

The cosmic-ray-normalized models \citep{wb1997,waxman2003,probation2011} assume protons emitted by GRBs are the dominant sources of the highest energy cosmic rays observed, and with these models we place limits on this assumption.  The neutrino spectral break and normalization depend on the bulk Lorentz boost factor $\Gamma$ and gamma-ray break energy in the GRB fireball.  Typical values required for pion production lead to the emission of $\sim 100$ TeV neutrinos.  In Section \ref{sec:results}, we place limits on the expected neutrino flux at different break energies.

The gamma-ray-normalized models \citep{winter2012,zhang2012} do not relate the observed cosmic ray flux to neutrinos produced in GRB fireballs, and with these models we place limits on internal fireball parameters.  We consider three types of gamma-ray-spectrum-normalized fireball models, calculated on a burst-by-burst basis, that differ in their neutrino emission sites.  The internal shock model relates the neutrino production radius to the variability time scale of the gamma-ray light curves \citep{wb1997,winter2012,zhang2012}.  The photospheric model places the radius at the photosphere through combinations of processes such as internal shocks, magnetic reconnection, and neutron-proton collisions \citep{rees2005,murase2008photo,zhang2011}.  The internal collision-induced magnetic reconnection and turbulence (ICMART) model favors a neutrino production radius $\sim 10$ times larger than the standard internal shock model due to a Poynting-flux-dominated outflow that remains undissipated until internal shocks destroy the ordered magnetic fields \citep{zhang2011,zhang2012}.  These models assert that gamma-ray emission and proton acceleration occur at the same radius.  This equivalence is not necessarily true for scenarios other than the single-zone internal shock model \citep{multipleis2015}, but allows the predicted neutrino flux to scale linearly with the proton-to-electron energy ratio in the fireball.  Additionally, the models addressed in this work do not account for a possible enhancement to the high energy neutrino flux due to acceleration of secondary particles \citep{klein2013muon,winter2014muon}.

We calculate the per-GRB predictions normalized to the measured gamma-ray spectra numerically with a wrapper of the Monte-Carlo generator SOPHIA \citep{sophia}, taking into account the full particle production chain and synchrotron losses in inelastic $p\gamma$ interactions.  For these calculations, we parametrize the reported gamma-ray spectrum of each GRB as a broken power-law approximation of the Band function \citep{band1993,zhang2012}.  We retrieve GRB parameters from the circulars published by satellite detectors on the Gamma-ray Coordinates Network\footnote{\url{http://gcn.gsfc.nasa.gov}} and the Fermi GBM database \citep{fermicat,fermispec}.  We compile the relevant temporal, spatial, and spectral parameters used in this analysis on our GRBweb database, presented on a publicly available website\footnote{\url{http://grbweb.icecube.wisc.edu}} \citep{grbwebicrc}.  The prompt photon emission time ($T_{100}$) is defined by the most inclusive start and end times ($T_1$ and $T_2$) reported by any satellite.  We use the most precise localization available.  Following the same prescription of our previous model limit calculations \citep{grbic40paper,ic59nature,grb4year}, if the fluence is unmeasured, we use an average value of $10^{-5}$ erg cm$^{-2}$; if the gamma-ray break energy is unmeasured, we use 200~keV for $T_{100} > 2$~s bursts and 1000~keV for $T_{100} \leq 2$~s bursts; and if the redshift is unmeasured, we use 2.15 for $T_{100} > 2$~s bursts and 0.5 for $T_{100} \leq 2$~s bursts.

The neutrino flux predictions depend on several unmeasured quantities; we use variability time scale 0.01 s and isotropic luminosity $10^{52}$ erg cm$^{-2}$ for long bursts and variability time scale 0.001 s and isotropic luminosity $10^{51}$ erg cm$^{-2}$ for short bursts, which are consistent with the literature \citep{baerwald2014,winter2012,zhang2012}.  If the redshift is known for a particular burst, we calculate the approximate isotropic luminosity from the redshift, photon fluence, and $T_{100}$ \citep{winter2012}.

\begin{figure}[h]
 \plotone{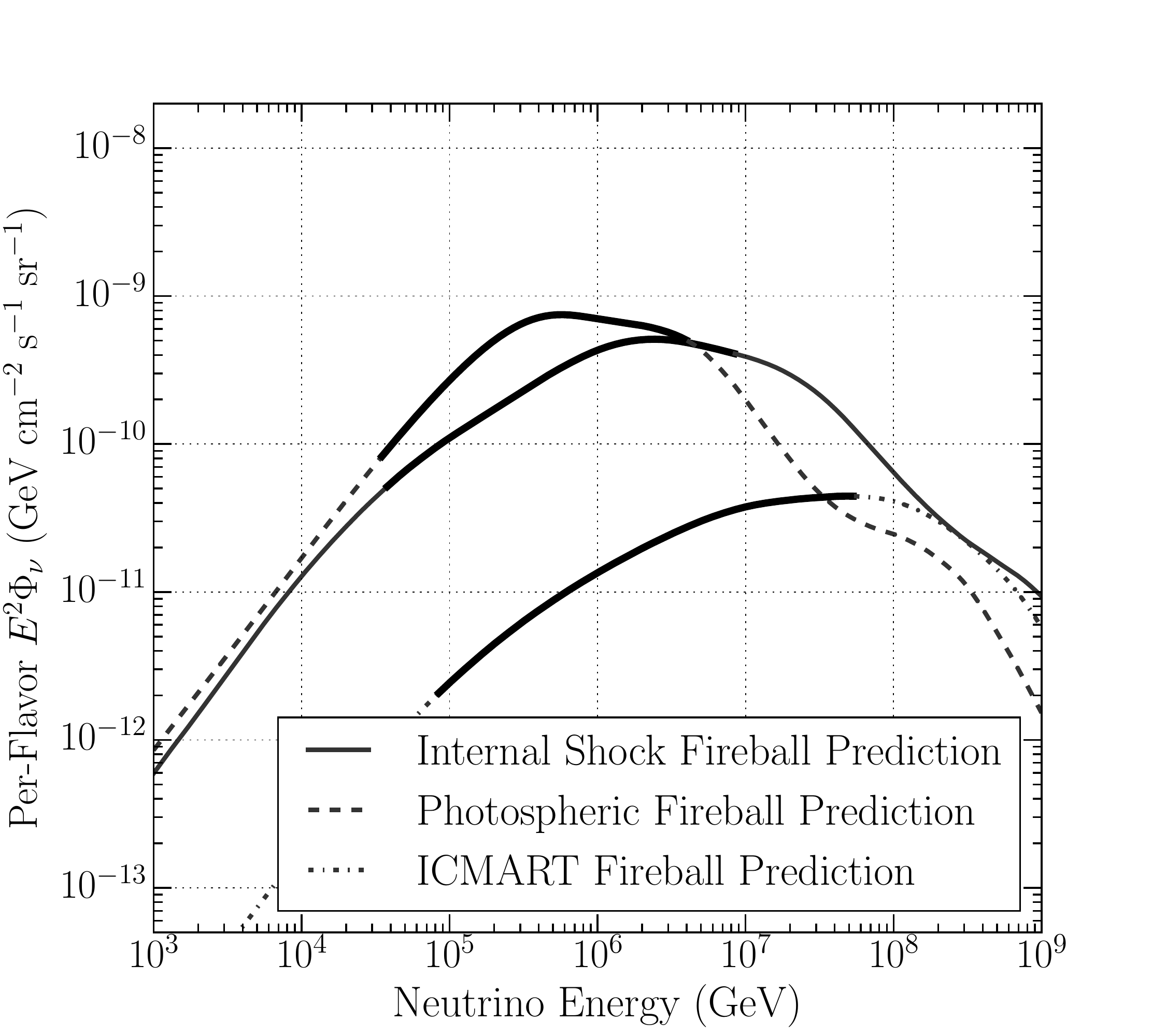}
 \caption{Per-flavor quasi-diffuse all-sky flux predictions, calculated with the $\gamma$ spectra of all GRBs included in this three-year search, for three different models of fireball neutrino production.  These fluxes assume $\Gamma=300$, $f_p=10$, full flavor mixing at earth, and 667 observable GRBs per year.  These models differ in the radius at which $p\gamma$ interactions occur.  The solid segments indicate the central 90\% energies of neutrinos that could be detected by IceCube. 
 \label{grbspectra}}
\end{figure}

Figure \ref{grbspectra} illustrates neutrino spectra from the three models with benchmark fireball parameters.  These benchmark parameters are bulk Lorentz boost factor $\Gamma = 300$ and proton-to-electron energy ratio, or baryonic loading, $f_p = 10$.  These spectra are presented as per-flavor quasi-diffuse fluxes, in which we divide the total fluence from all GRBs in the sample by the full sky $4\pi$ steradians and one year in seconds, and scale the total number of bursts to a predicted average 667 observable bursts per year, which has been used in our previous IceCube publications \citep{ic22paper,grbic40paper,ic59nature,grb4year}.  The actual number of bursts observed by satellite detectors in each year is less than the predicted average because of detector field of view limitations and obstruction by the sun, earth, and moon.  For all GRBs we assume that neutrino oscillations over cosmic baselines result in equal fluxes for all flavors \citep{aartsen2015flavor, aartsen2015combined, palomares2015spectral, bustamante2015theoretically, palladino2015flavor}.

\section{The IceCube Detector}
\label{sec:detector}

The IceCube detector \citep{icfirstyear} consists of 5160 digital optical modules (DOMs) instrumented over 1 km$^3$ of clear glacial ice 1450 to 2450~m below the surface at the geographic South Pole.  IceCube is the largest neutrino detector in operation.  Each DOM incorporates a 10~in. diameter photomultiplier tube (PMT) \citep{pmtpaper}.  Signal and power connections between the DOMs and the computer farm at the surface are provided by 86 vertical in-ice cables or ``strings" that each connect 60 DOMs spaced uniformly.  Adjacent strings are separated by about 125 m.  The DeepCore array \citep{dcdesignperf} is made up of a more densely spaced subset of strings that are located in the clearest ice at depths below 2100~m and contain higher quantum efficiency PMTs.

Sensor deployment began during the 2004-2005 austral summer. Physics data collection began in 2006 with the 9-string configuration and continued with partial detector configurations through completion of the 86 strings in December 2010.  We conducted the analysis presented in this paper using data taken from May 2010 through May 2013, with one year using the 79-string configuration and two years using the completed detector.

The DOM PMTs detect the Cherenkov radiation of relativistic charged particles produced in deep inelastic neutrino-nucleon scattering in or near the detector volume.  Data acquisition \citep{daqpaper} begins once the output current exceeds the threshold of $1/4$ of the mean peak current of the pulse amplified from a single photo-electron.  If another such ``hit" is recorded in a neighboring or next-to-neighboring DOM within 1 $\mu$s, the full waveform information is recorded.  DOMs with hits failing this local coincidence condition report a short summary of their recorded waveform for inclusion in data records.  The digitized waveforms are sent to the computers at the surface, where they are assembled into ``events" by a software trigger \citep{daqpaper,kelley2014trigger}.  The relative timing resolution of photons within an event is about 2~ns \citep{icfirstyear}.  Data are sent daily from the detector to computers in the Northern Hemisphere via satellite.

There are two main interaction topologies that can manifest in hit DOMs during an event: a thin track or a near-spherical shower.  Track-like light emission is generated by muons created either through cosmic ray interactions in the atmosphere or $\nu_{\mu}$ interacting in the ice.  Shower-like light emission, which is the signal of this analysis, is generated by electromagnetic cascades from charged current interactions of $\nu_e$ and $\nu_{\tau}$ and hadronic cascades from neutral current interactions of $\nu_e$, $\nu_{\tau}$, and $\nu_{\mu}$ in the ice.  $\nu_{\alpha}$ stands for $\nu_{\alpha} + \bar{\nu}_{\alpha}$ unless otherwise specified.

Electromagnetic cascades of photons, electrons, and positrons develop through radiative processes, such as bremsstrahlung and pair production.  Once the energy of each particle in the cascade is below the critical energy, non-radiative processes, such as ionization and Compton scattering dominate and the shower stops growing.  Hadronic cascades result from showers of baryons and mesons produced by deep inelastic scattering interactions of all neutrino types.  Interactions of $\bar{\nu}_e$ with electrons via the Glashow resonance at 6.3~PeV would also produce cascade signal for this analysis; however, this resonance has not yet been observed.

\section{Simulated Data and Reconstruction}
\label{sec:simandreco}

We use Monte Carlo simulations of neutrinos interacting in the IceCube detector for the signal hypothesis in our event selection and optimization for this search.  Neutrinos are simulated with the \texttt{NEUTRINO-GENERATOR} program, a port of the \texttt{ANIS} code \citep{nugen2005}.  We use \texttt{NEUTRINO-GENERATOR} to distribute neutrinos with a power-law spectrum uniformly over the entire sky and then we propagate them through the earth and ice.  The simulated neutrino-nucleon interactions take cross sections from CTEQ5 \citep{cteq5}.  The Earth's density profile is modeled with the Preliminary Reference Earth Model \citep{prem1981}.  The propagation code takes into account absorption, scattering, and neutral-current regeneration.

Although we use data outside of GRB gamma-ray emission time windows for our background, simulated neutrinos and muons generated in cosmic ray air showers are useful checks to characterize our background and estimate the signal purity of our final data sample.  We apply the Honda et al. spectrum \citep{honda2006} for the atmospheric neutrino background.  We use the  \texttt{CORSIKA} simulation package \citep{heck1998} to simulate cosmic ray air showers.  Muons are traced through the ice and bedrock incorporating continuous and stochastic energy losses \citep{mmc}.  The PMT detection of Cherenkov light from muon tracks and showers is simulated using ice and dust layer properties determined in detailed studies and simulations \citep{photonics} \citep{ice}.  Finally, we simulate the DOM triggering and signal from all interactions.  We process these signals in the same way as described in Section \ref{sec:detector}.

We use the geometric pattern, timing, and amount of recorded photons in an interaction to reconstruct and identify the incident particle.  We run simple analytic reconstructions automatically in data filters at the Pole and more complex reconstructions on the reduced data in the north.  These reconstructions provide powerful discriminating features that we employ in our event selection methods, described in the next section.

One of the analytic calculations used to identify shower-like events is the ``tensor-of-inertia" algorithm of \cite{amandareco} that determines an interaction vertex by calculating a ``center of mass," where the mass terms correspond to the number of photoelectrons recorded by each DOM.  The eigenvector along the longest principle axis provides a reasonable guess for the incident particle trajectory.  Using the rigid body analogy, a spherical shower should provide nearly even tensor-of-inertia eigenvalues, while an elongated track should have one eigenvalue much smaller than the other two.

A second analytic calculation is the ``line-fit" algorithm of \cite{amandareco} that ignores the geometry of the Cherenkov cone and the optical properties of the medium and assumes light travels along a one-dimensional path at some constant velocity through the detector.  This algorithm determines a best-fit track, but provides a good indicator for spherical showers, which exhibit slower best-fit velocities than tracks.  We recalculate this fit during the final level of event selection using an improved version of the algorithm that discards photons due to uncorrelated noise \citep{improvedlf}.

We also perform fast reconstructions based on an analytic likelihood function.  In general, we determine the set of parameters $\textbf{a}=\left\{a_1,a_2,...,a_m\right\}$ of an event that maximizes the likelihood function

\begin{equation}
\mathcal{L}(\overrightarrow{\textbf{x}} \vert \textbf{a}) = 
\prod_{i=1}^{N} p(\textbf{x}_i \vert \textbf{a})
\end{equation}
given sets of observables $\overrightarrow{\textbf{x}}=\left\{\textbf{x}_1,\textbf{x}_2,...,\textbf{x}_N\right\}$ from N DOMs.  We minimize the negative logarithm of $\mathcal{L}$ with respect to $\textbf{a}$.  In these online likelihood reconstructions, the $\overrightarrow{\textbf{x}}$ are parametrized in terms of the difference between the recorded photon hit times and the expectation without scattering or absorption \citep{amandareco}.

For a cascade hypothesis, which can be considered point-like because of the relatively small maximum shower range on the scale of IceCube instrumentation \citep{ereco2014}, the time residual depends on the location and time of the interaction vertex $\textbf{a} = \left\{t,x,y,z\right\}$.  This minimization is seeded by the above center of mass vertex calculation and the earliest vertex time for which at least 4 DOMs record hits with time residuals within [0, 200] ns.  For an infinite track hypothesis, the time residual depends on the distance from an arbitrary point on the track and the direction of the track $\textbf{a} = \left\{x,y,z,\theta,\phi\right\}$.  This minimization is seeded by the above line-fit track.  Both of these cascade and track hypothesis maximum likelihoods provide useful best-fit quantities for the event selection described in Section \ref{sec:eventselect}.

The more computationally expensive likelihood-based reconstruction that we use on the reduced data takes advantage of detailed in-ice photon propagation models \citep{photospline,photonics} and DOM waveform information.  The detection of light follows a Poisson distribution and thus the likelihood function we maximize is

\begin{equation}
\mathcal{L}(k \vert E,t,x,y,z,\theta,\phi) = \prod_{i=1}^{N_{DOM}} \frac{\lambda_{i}^{k_i}}{k_{i}!} \e^{-\lambda_i}
\label{mpodllh}
\end{equation}
for the observed and expected number of photoelectrons $k$ and $\lambda$ in each DOM from a neutrino traveling in direction $(\theta$,$\phi)$ and depositing energy $E$ at time $t$ and vertex $(x,y,z)$ \citep{ereco2014}.  This likelihood in general considers multiple light sources, e.g. for stochastic muon losses along a path through the detector.  Because we define our signal to be neutrino-induced showers, it is sufficient to use a point-like hypothesis.  Due to the linear relationship between Cherenkov light emission and deposited energy in $\nu_e$ charged-current interactions, we estimate the electromagnetic-equivalent energy of an event by comparing to a template electromagnetic cascade of 1 GeV \citep{ereco2014}.  By using Cherenkov light source-dependent spline-fit tables of photon amplitudes and time delays, we minimize \citep{minuit} the negative log-likelihood and extract the best-fit parameters of the particle that produced the shower.

\begin{figure}[h]
 \plotthree{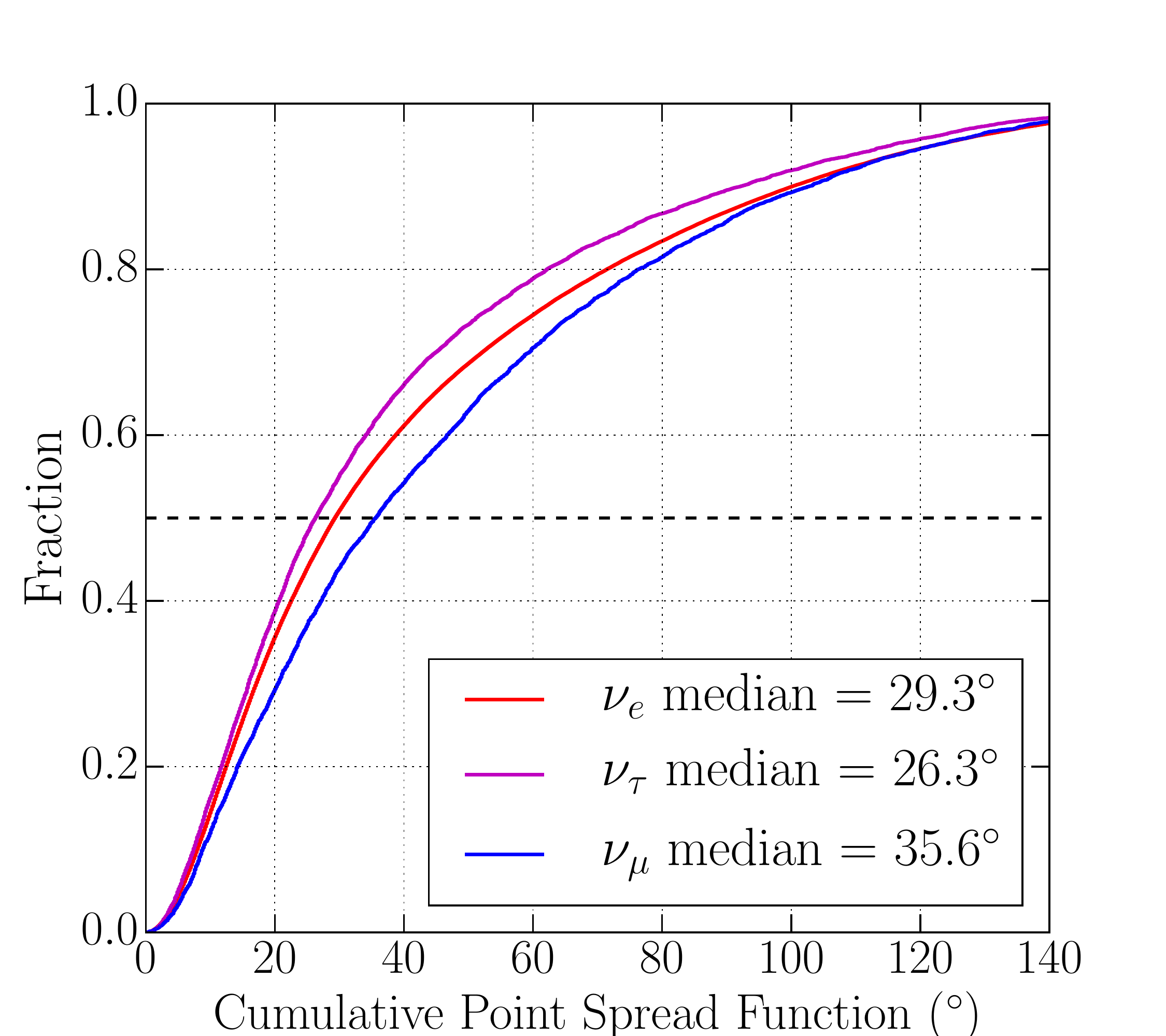}{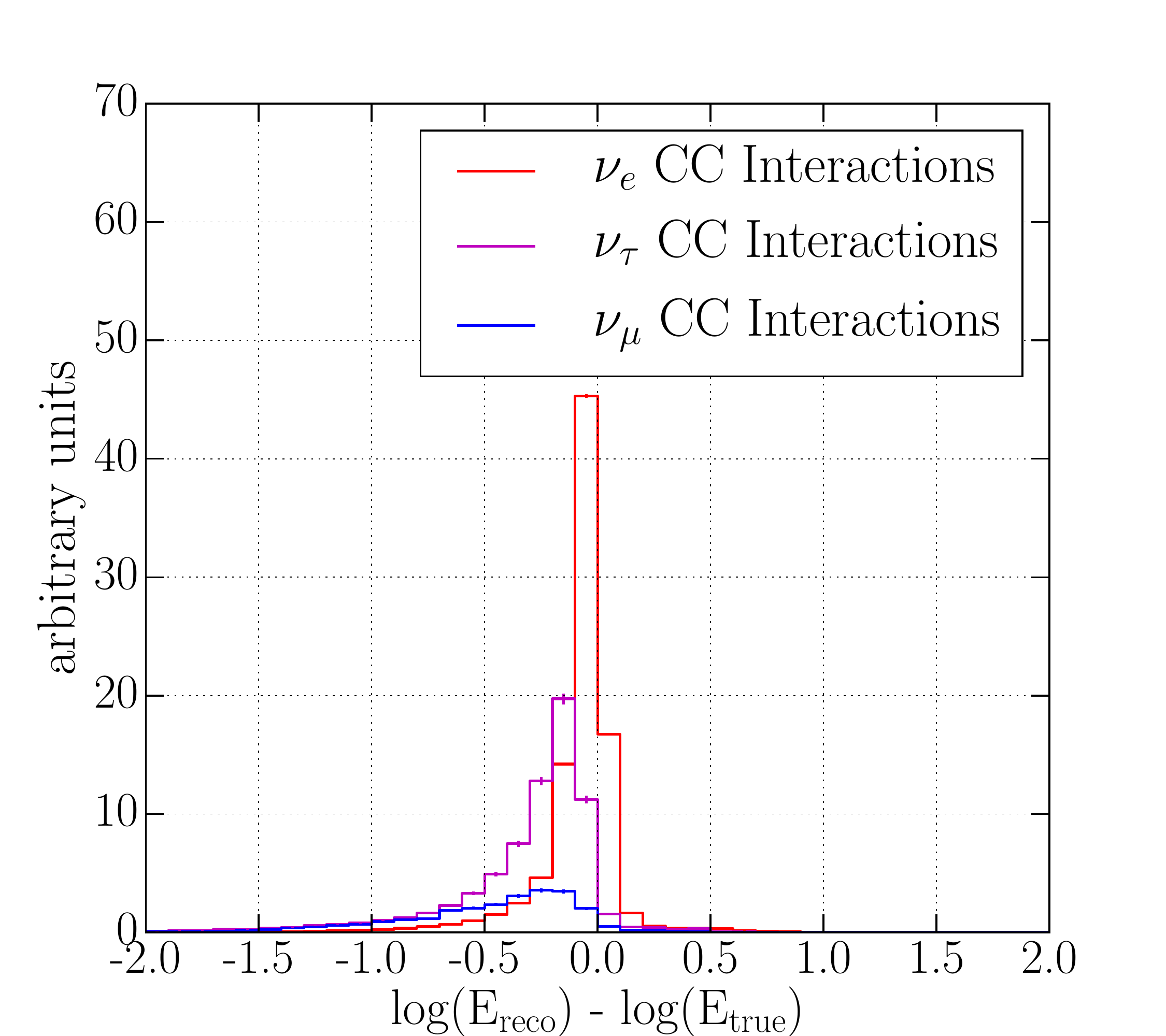}{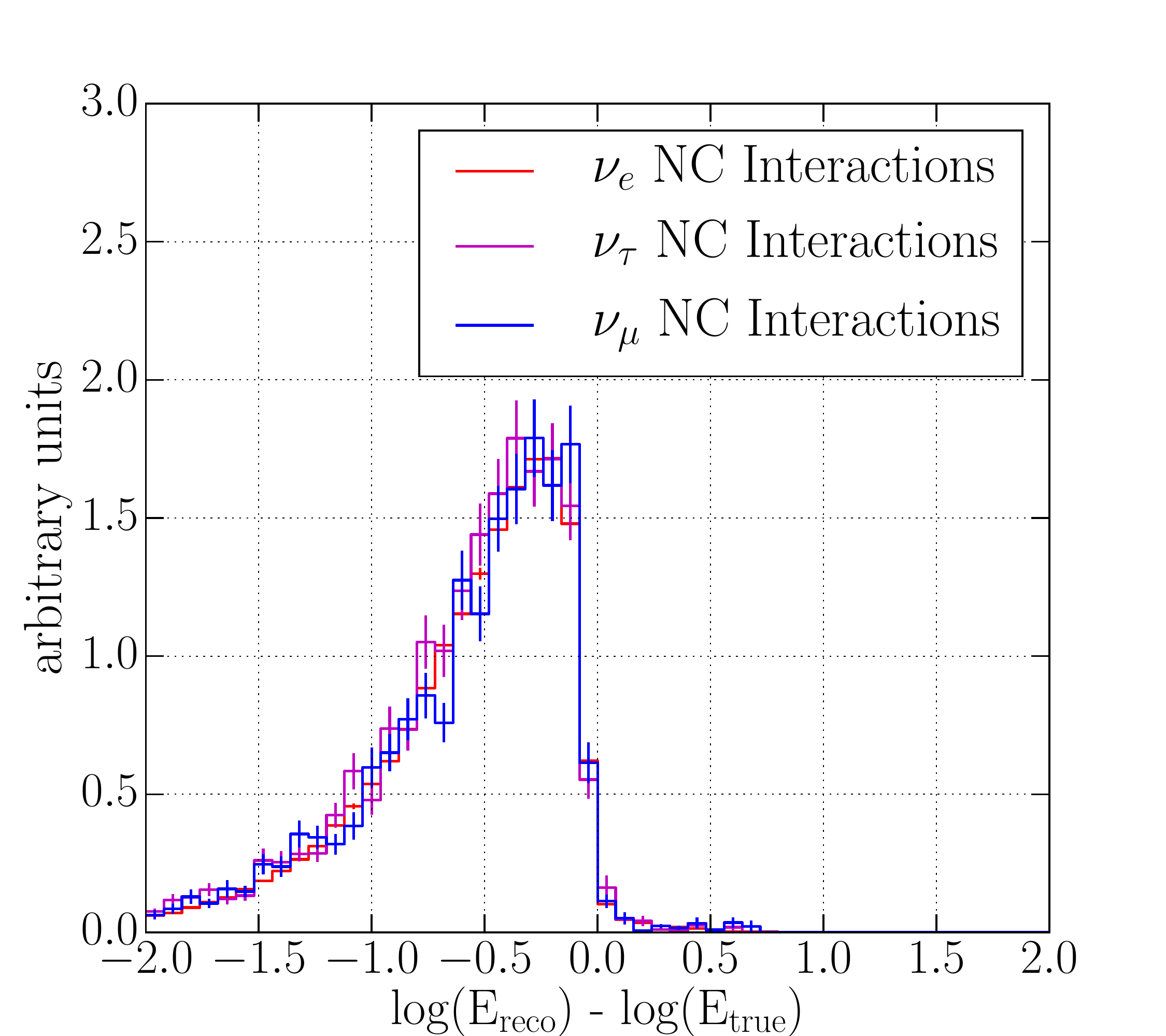}
 \caption{Left: cumulative point spread function per neutrino flavor at final cascade event selection. Middle: energy resolution per neutrino flavor for charged-current interactions at final event selection. Right: energy resolution per neutrino flavor for neutral-current interactions at final event selection.
 \label{angres-eres}}
\end{figure}

We iterate this minimization several times and achieve improved angular reconstruction from the best-fit energy and modeled ice structure.  The reconstruction is seeded by the tensor-of-inertia direction, online likelihood vertex, and best-fit energy from a simple isotropic shower and coarsely binned propagation model expectation in Equation \ref{mpodllh}.  We use this reconstruction's best-fit direction and energy for each event in our likelihood analysis.  We estimate a lower bound on the error in the reconstructed directions with the Cramer-Rao relation \citep{cramer,rao} between the covariance of each fit parameter and the inverted Fisher information matrix.  Figure \ref{angres-eres} shows the angular and energy resolutions using this reconstruction at the final event selection of this cascade search.  The reconstructed energy for neutral current interactions is less than the primary neutrino energy because of the dissipation to outlets without Cherenkov emission.

\section{Event Selection}
\label{sec:eventselect}

Our signal in this search is one or more high energy neutrino-induced showers coincident in space and time with one or more GRBs.  Before we analyze the likelihood that an event is a GRB-emitted neutrino, we must reduce the triggered data that is dominated by nearly 3 kHz of cosmic ray air shower muons to a much smaller selection of possible signal.  Our event selection is optimized on a general $E^{-2}$ spectrum of diffuse astrophysical simulated $\nu_e$ for signal and off-time data for background; we choose ``off-time" as data not within two hours of any reported GRB $\gamma$ emission in order to ensure blindness to potential signal. We use the reconstructions detailed in Section \ref{sec:simandreco} to remove the background and realize our final sample.

Searches for neutrino-induced showers from astrophysical and atmospheric sources have been conducted previously in IceCube \citep{firstcascade,ic40cascade,hese2014,mese2015}.  The predominant difference between these and the search presented in this paper is that we assume neutrinos come from known transient sources.  The previous shower-like event selections assume a diffuse signal or constantly emitting sources and, as a result, require much more stringent background reduction that leads to data rates nearly a factor of 100 smaller than what is needed in this search.  Cascade containment constraints in the detector were imposed to reach these low backgrounds, which are achieved in this search by the effective cuts in time and space around each GRB in the unbinned likelihood analysis presented in Section \ref{sec:llh}.

The first class of background to remove is track-like events generated by muons losing their energy through continuous ionization processes.  These muon tracks are relatively easily separated from our shower-like neutrino signal by use of a filter run online at the detector site.  During the 79-string configuration, we use the tensor-of-inertia eigenvalue ratio and line-fit absolute speed to choose spherical events.  The online cascade filter for the first two years of the 86-string configuration imposes a zenith-dependent selection to allow more events with energies below 10 TeV.  We use the reduced log-likelihood calculated in the fast cascade likelihood reconstruction to remove a large fraction of the muons that are misreconstructed as upgoing.  The down-going region requires more restrictive reduced log-likelihood values in combination with the same online 79-string event selection from above.

Upon transmission via satellite from the South Pole, these $\sim 30$~Hz filtered data of spherically shaped events must be further reduced to have optimal performance with the boosted decision tree (BDT) forest algorithm described below.  The background at this rate is still dominated by atmospheric muons that lose their energy through bremsstrahlung, photonuclear interactions, and pair production.  These stochastic energy loss mechanisms create nearly spherical hit patterns that are difficult to differentiate from our signal when the muon track is at the edge or outside of the detector and therefore not observed.  We derive variables from the more CPU-intensive reconstructions run offline to separate these muons from neutrino-induced showers at this selection level, called level 3 in this text, and for eventual machine-learning input.

Even though the background muons in the online filter event selection are shower-like in topology, we exploit the Cherenkov light hit patterns of the minimally ionizing muon track when possible to differentiate from showers produced by neutrinos.  We perform this further separation of signal from muon background first by using the fast track hypothesis and cascade hypothesis analytic likelihood function reconstructions, described in Section \ref{sec:simandreco}.  Our first level 3 selection is on events for which the ratio of the track likelihood to the cascade likelihood heavily favors the cascade hypothesis.  Two other background muon features that we take advantage of are their down-going directions and typically lower energies.  We thus parametrize the track likelihood reconstructed zenith in terms of reconstructed energy and remove low energy down-going events from our sample.  Lastly at this rate, we choose events based on the fraction of hit DOMs that are inside of a sphere centered on the reconstructed vertex with a radius determined by the mean hit distance.  We optimize the radius and fraction filled of the sphere for reconstructed vertices inside and outside of the instrumented volume and use both volume regions for this neutrino search.

Our final event selection employs a machine learning algorithm to optimize the separation of signal and background over a space of many features.  The algorithm we use is a BDT forest \citep{freundbdt1997} that also has been used in Northern Hemisphere $\nu_{\mu}$ track GRB coincidence searches in IceCube \citep{ic59nature,grb4year}.  We built a collection of variables, many of which were influenced by past neutrino-induced shower searches conducted in IceCube \citep{firstcascade,ic40cascade}, to train the BDT to discriminate between signal and background events.  During this training, the algorithm determines the variable value that best separates signal and background at each node of each tree.  Each successive tree increases, or ``boosts," the weights of incorrectly classified events from the prior tree, allowing more ambiguous events to be classified correctly.  The trees and forest grow until certain stopping criteria are reached.  The signal/background $(+1/-1)$ score of each event that we use in our final selection is a weighted average of its scores over all trees, where this weight is related to the boost factor of each tree.  For each of the three year-long detector configurations of this search, we train a separate BDT with configuration-specific signal simulation and background data.

\begin{figure}[h]
  \plotthree{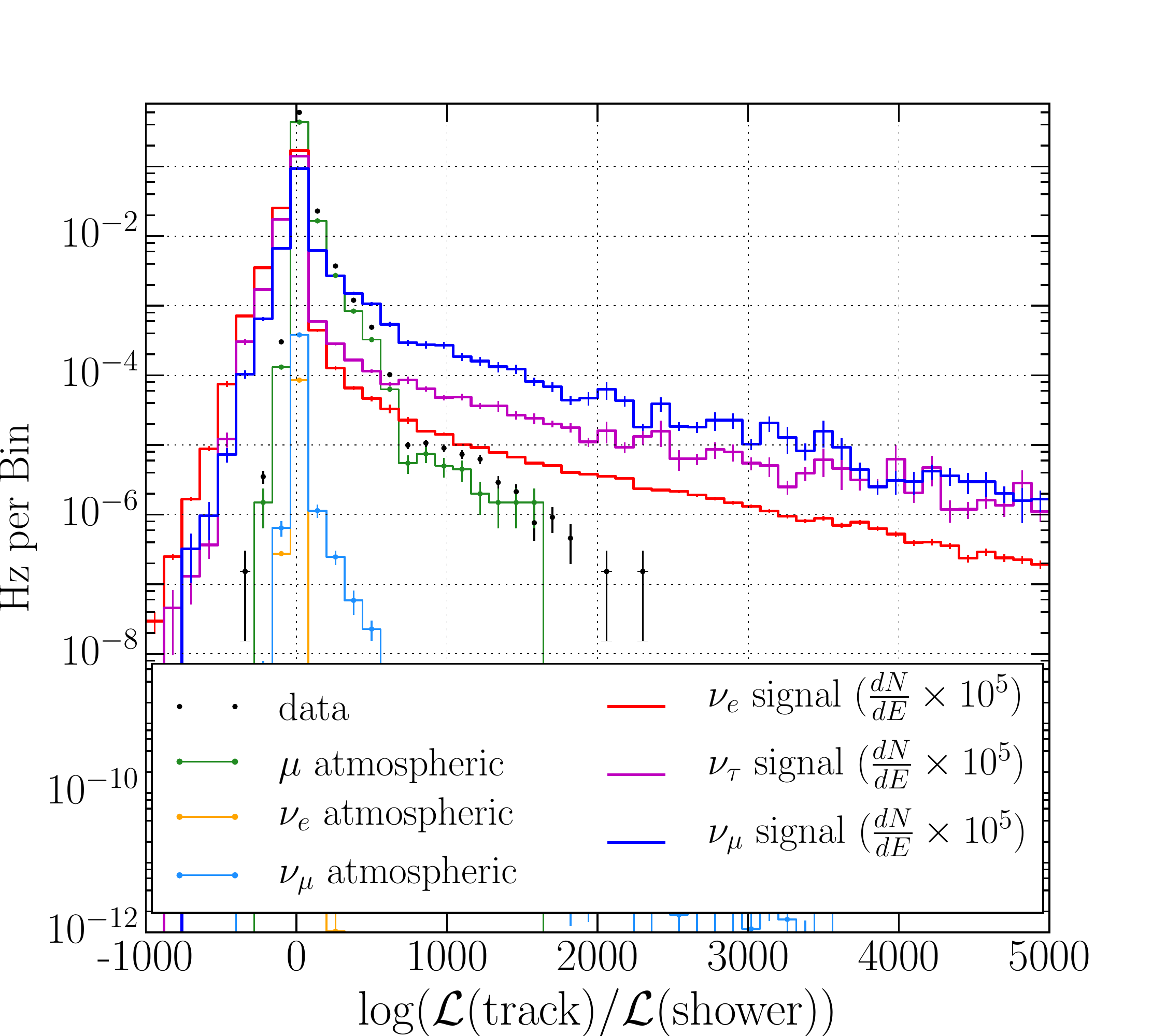}{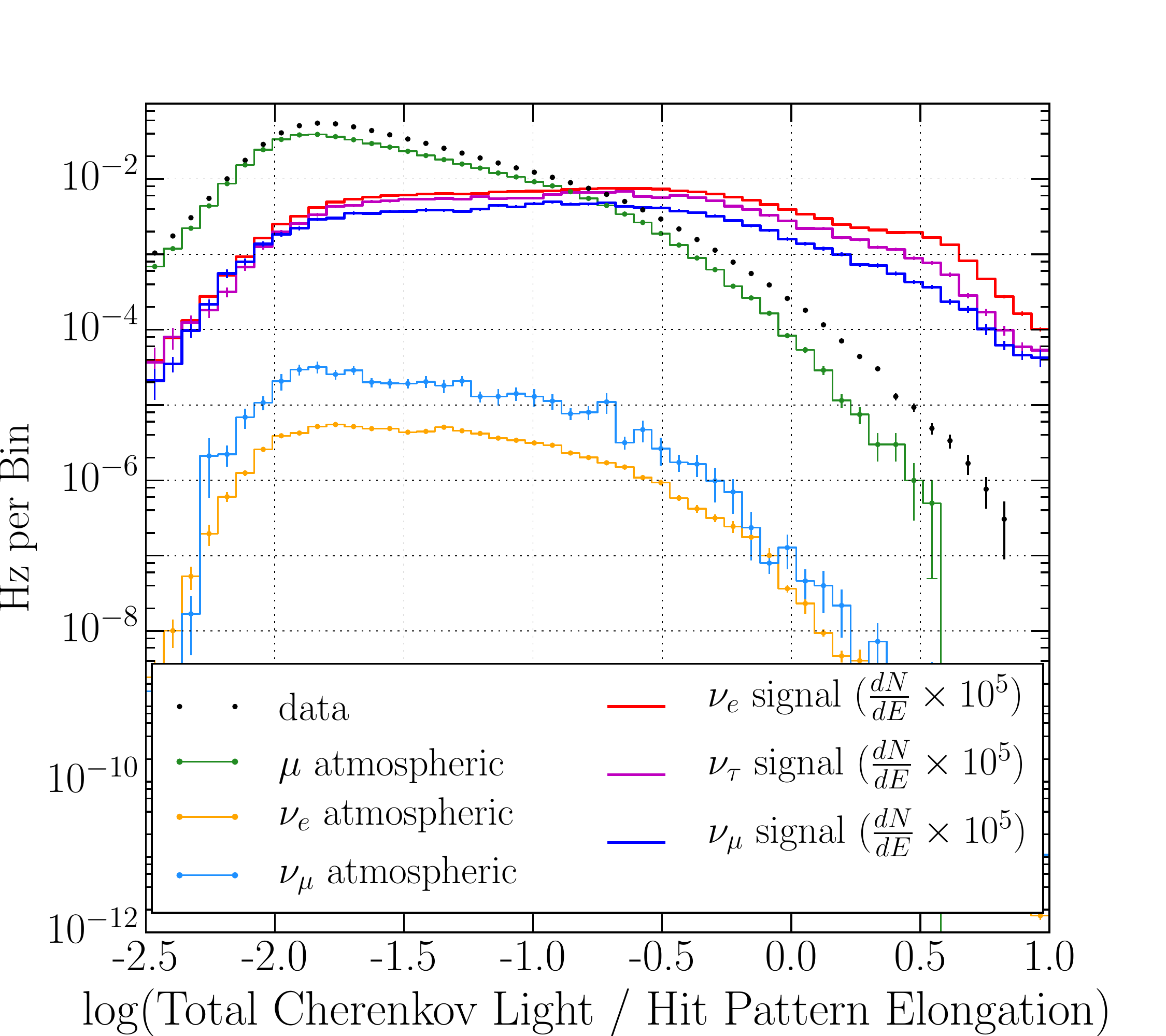}{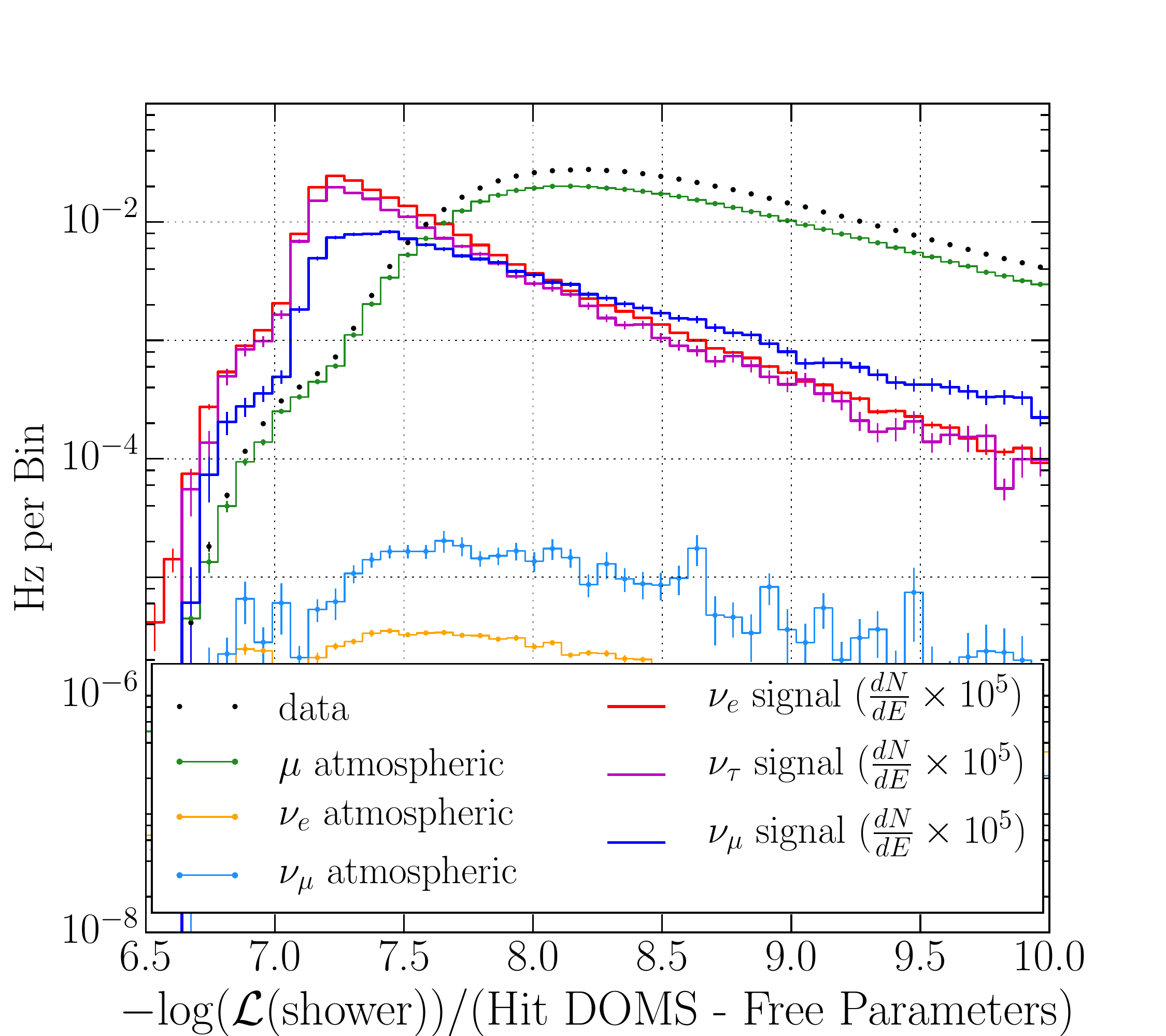}
 \caption{Distributions for three of the most powerful signal/background discrimination variables used by the BDT forest, shown after the level 3 event selection.  $E^{-2}$ $\nu_e$ (red solid line) was used for signal and data (black dots) was used for background in the BDT forest training.  $\frac{dN}{dE}$ is defined in Figure \ref{bdtdist}. 
 \label{vars-level3}}
\end{figure}

\begin{figure}[h]
  \plotthree{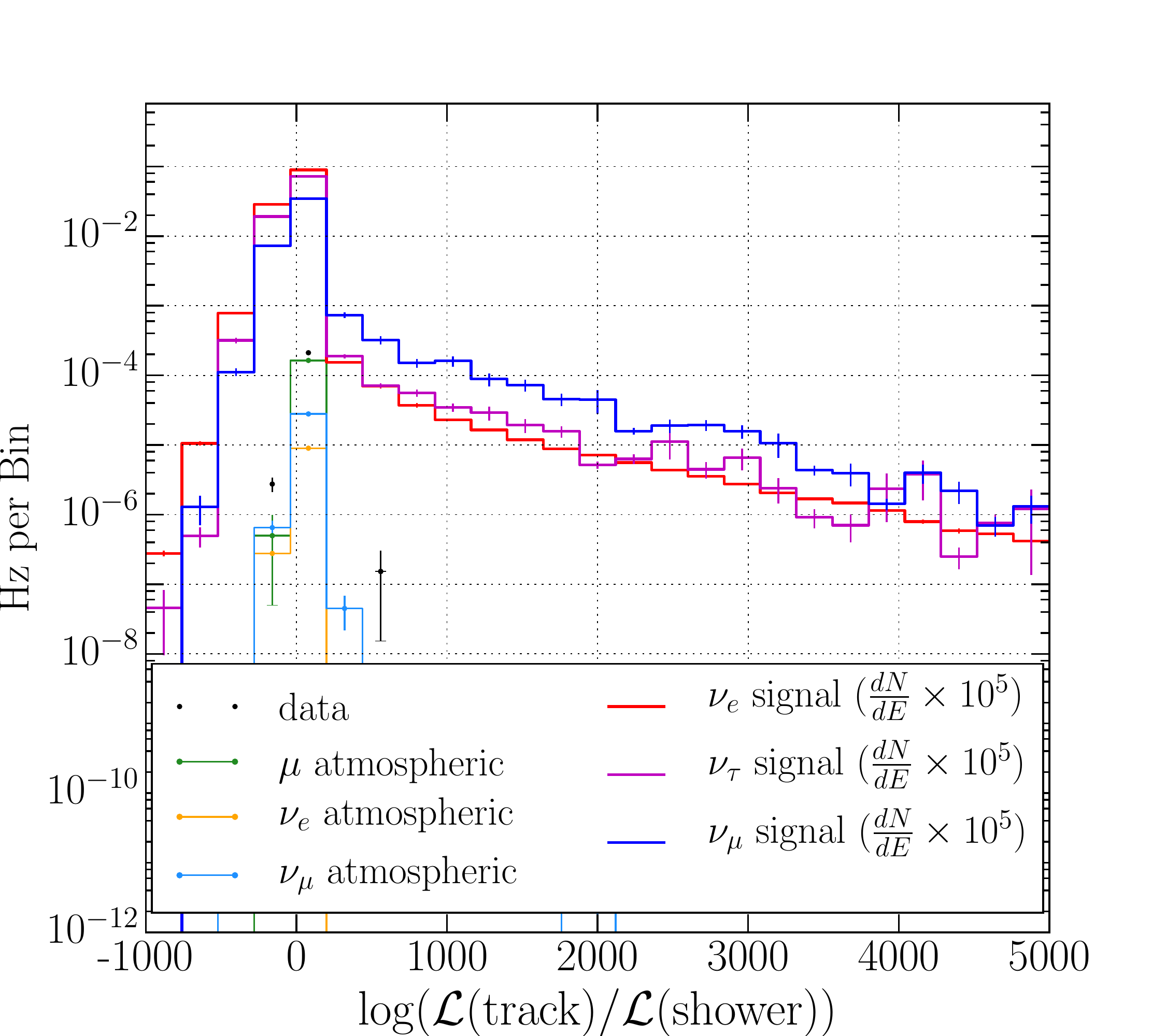}{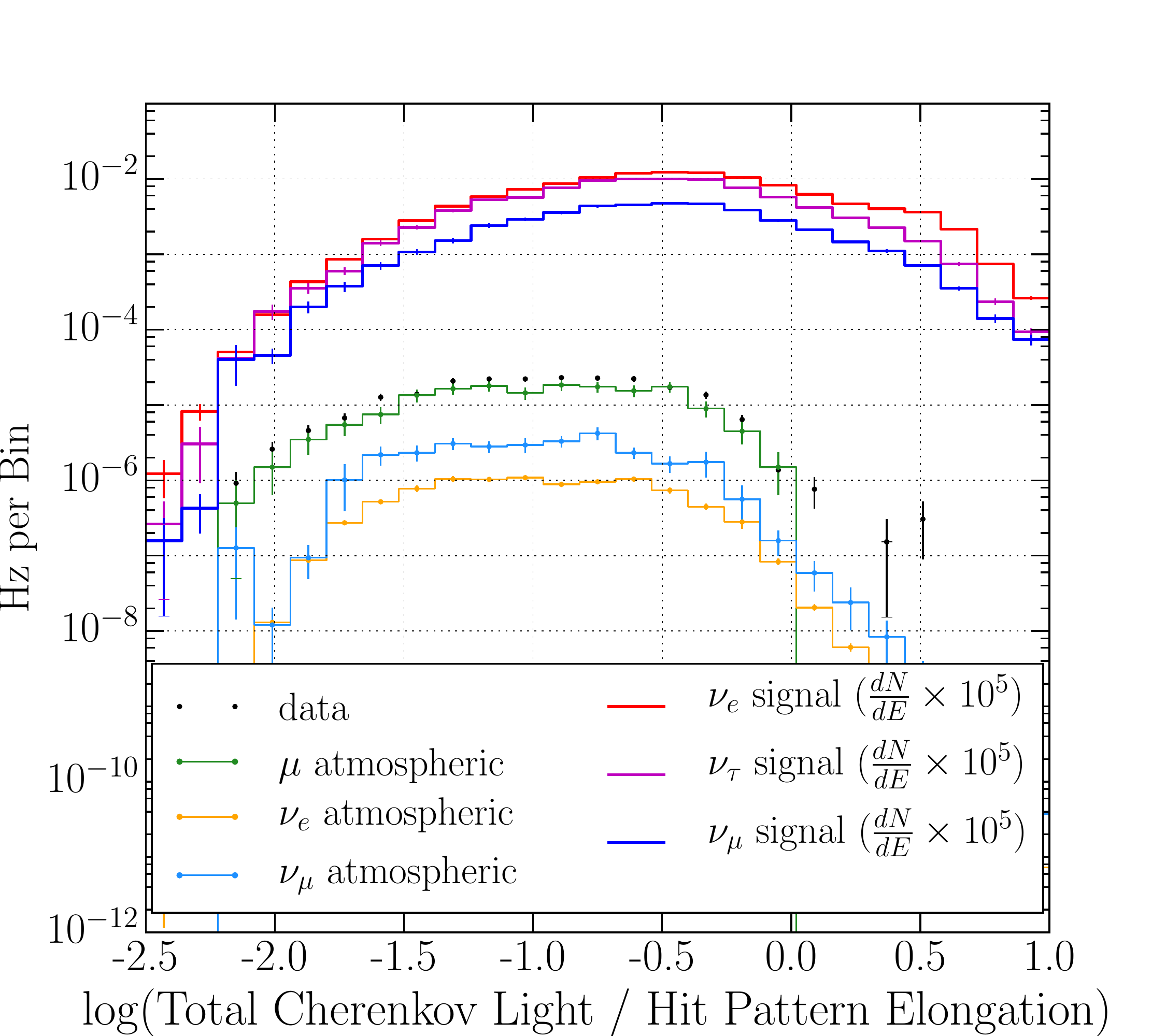}{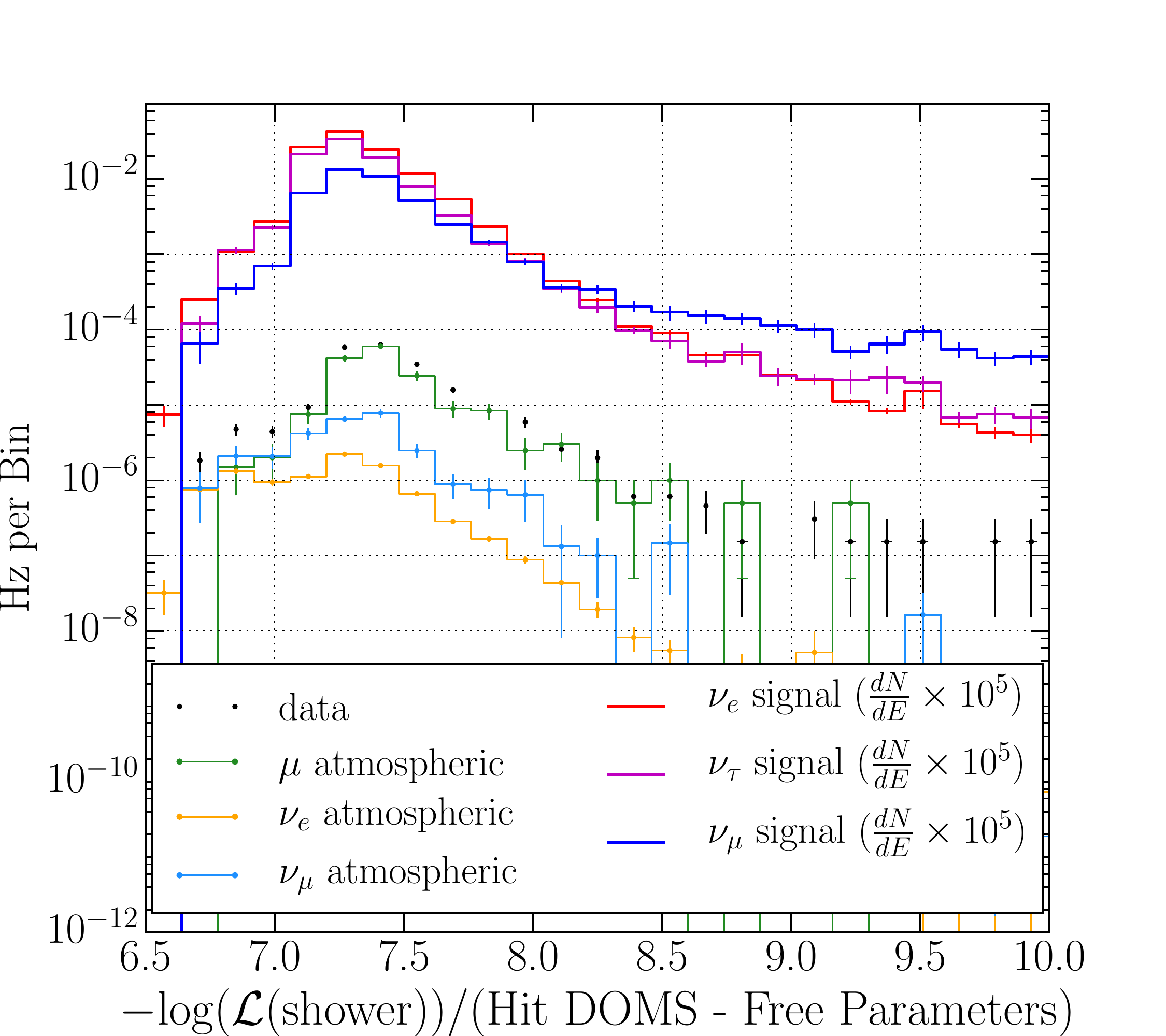}
 \caption{Distributions for three of the most powerful signal/background discrimination variables used by the BDT forest, shown after the final event selection.  $E^{-2}$ $\nu_e$ (red solid line) was used for signal and data (black dots) was used for background in the BDT forest training.  $\frac{dN}{dE}$ is defined in Figure \ref{bdtdist}. 
 \label{vars-final}}
\end{figure}

Our BDT discrimination variables take advantage of topological and energetic differences between astrophysical neutrino and atmospheric muon spectra.  Figures \ref{vars-level3} and \ref{vars-final} show the distributions for simulated astrophysical neutrinos, atmospheric neutrinos, atmospheric muons, and data for three of the most powerful variables used by the BDT after the level 3 and final level, respectively.  Some deviations in variable values between measured and simulated data are visible in Figures \ref{vars-level3} and \ref{vars-final}, and are attributed to limitations of simulation in reproducing all observed events.  These deviations do not affect the analysis because the background is determined from measured data.

\begin{figure}[h]
 \plotone{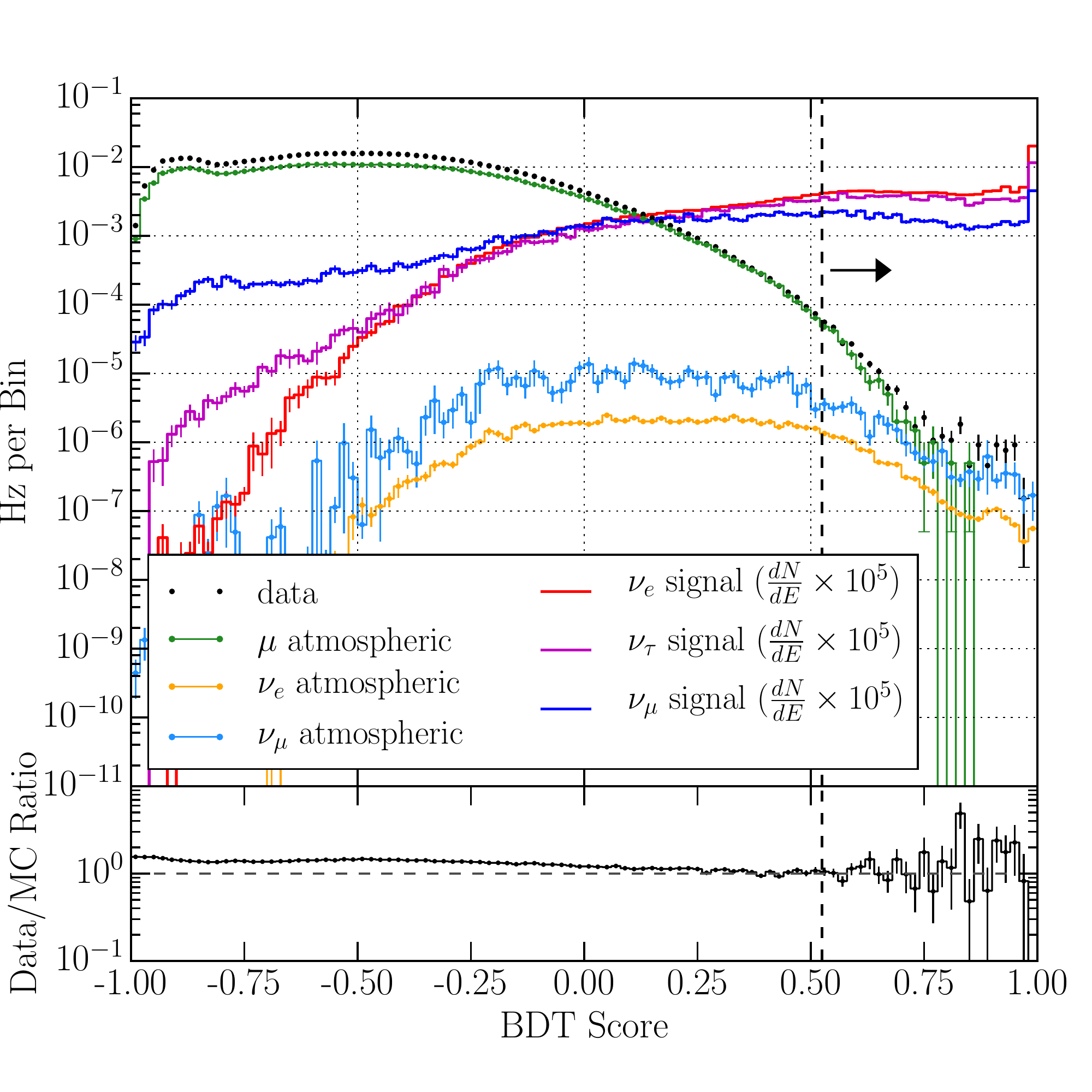}
 \caption{Distribution of data, simulated muon background, simulated atmospheric neutrino background, simulated $\mathrm{E}^{-2}$ neutrino signal (where $\frac{dN}{dE}=10^{-8}\mathrm{GeV}^{-1}\mathrm{cm}^{-2}\mathrm{s}^{-1}\mathrm{sr}^{-1}(\frac{E}{\mathrm{GeV}})^{-2}$) with respect to BDT score.  The vertical dashed line represents the final event selection of $> 0.525$.
 \label{bdtdist}}
\end{figure}

\begin{figure}[h]
 \plotone{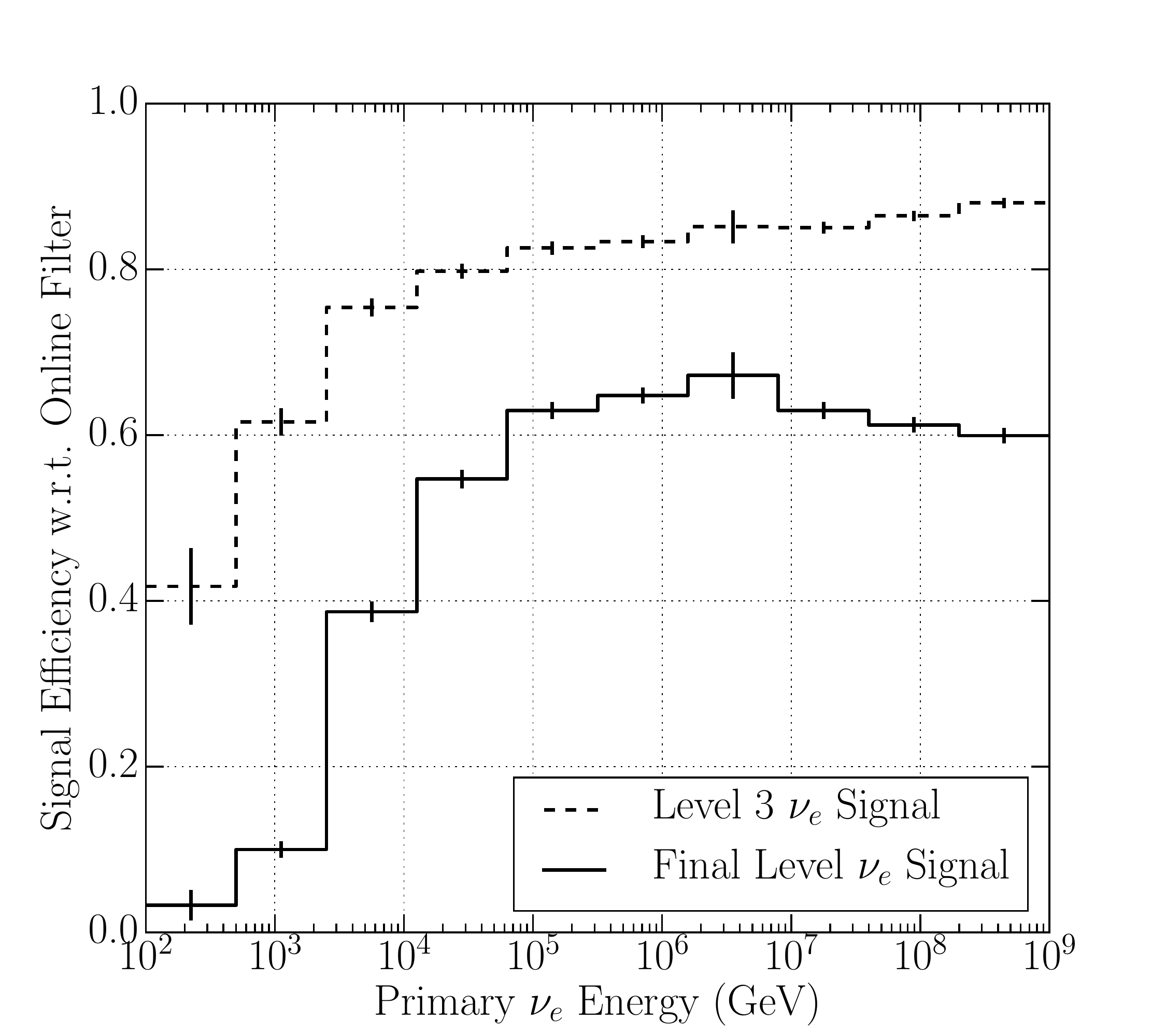}
 \caption{Signal efficiency relative to online filter level as a function of neutrino energy.  Signal is $E^{-2}$ $\nu_e$ simulation. Averaged over the three search years, the online filter level data rate is 30 Hz and is reduced to $1.6 \times 10^{-4}$ Hz at the final analysis level.
 \label{effs}}
\end{figure}

Figure \ref{bdtdist} shows the distributions of simulation and data with respect to BDT score.  The vertical dashed line corresponds to our optimized final event selection described in the next section.  The atmospheric and astrophysical $\nu_{\mu}$ in these plots is already preselected to be shower-like and has minimal overlap with the track search.  Background is reduced by a factor of $6 \times 10^{-6}$ from online filter level to the BDT-based final selection.  Figure \ref{effs} shows the $\nu_e$ signal efficiency at this final level with respect to the online filter rate as a function of energy.  The slight decrease in efficiency for the highest energy signal is due to events for which the computationally expensive reconstruction, described at the end of Section \ref{sec:simandreco} and used in the likelihood analysis, fails to converge.  These simulated neutrinos interact outside of the detector, account for about 1\% of the $E^{-2}$ signal, and are discarded prior to the BDT selection.

One of the most effective BDT variables is the same ratio of the track likelihood to the cascade likelihood used in level 3.  The high energy background muons that passed in spite of the selection at level 3 are distinguished from signal with the BDT algorithm.  The large likelihood ratios, seen in Figures \ref{vars-level3} and \ref{vars-final}, are due to the highest energy signal events with the most hit DOMs.  Another potent variable is the total amount of Cherenkov light imparted in the DOMs divided by the tensor-of-inertia derived elongation of an event.  The numerator separates lower energy background atmospheric muons from higher energy astrophysical signal neutrinos, while the denominator separates elongated background atmospheric muons from spherical neutrino-induced showers.  As shown in Figure \ref{vars-level3}, the ratio of these two observables effectively separates the lower energy atmospheric neutrino background from our signal as well.  A third useful variable is the reduced negative log-likelihood value from the analytic cascade hypothesis reconstruction.  

\begin{figure}[h]
 \plotone{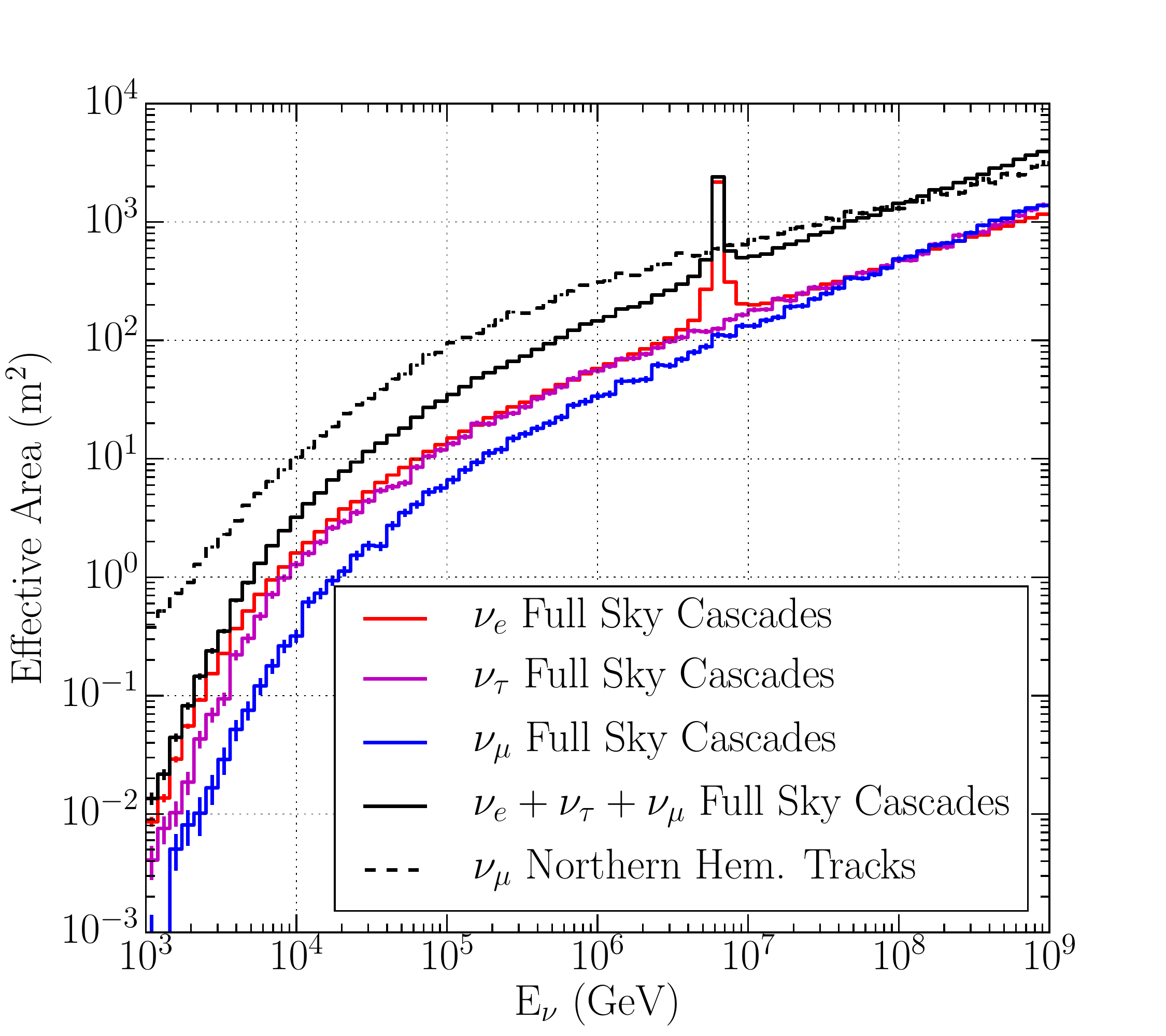}
 \caption{Effective areas for the full-sky shower-like and Northern Hemisphere track-like GRB-coincident event searches with the 79-string detector.  79- and 86-string detector effective areas are similar.
 \label{aeff}}
\end{figure}

Compared to the event samples of IceCube's searches for $\nu_{\mu}$-induced tracks from Northern Hemisphere GRBs, our backgrounds in this all-sky all-flavor cascade search require an event selection to a 10 times lower data rate of 0.16~mHz averaged over the three search years.  The differences between neutrino-induced showers and muon-induced stochastic energy loss showers are less apparent than the differences between neutrino-induced tracks and detector-edge and coincident muon-induced tracks incorrectly reconstructed as upgoing.  In the Northern Hemisphere track searches, most of these muons are able to be identified and removed, leaving only atmospheric neutrinos.  As a result, the atmospheric neutrino purity with respect to muons of this all-sky all-flavor search is $\sim$40\%, and is significantly less than the $\sim$90\% purity of the Northern Hemisphere track searches.  Nevertheless, we achieve similar sensitivity to the track search through our acceptance of $\nu_e$, $\nu_{\tau}$, and $\nu_{\mu}$ signal from GRBs over the entire sky.  Figure \ref{aeff} compares the effective areas for the two different GRB neutrino searches.

\section{Unbinned Likelihood Analysis}
\label{sec:llh}

Once we have selected a sample of events that resemble high energy neutrino-induced electromagnetic or hadronic showers, we must calculate the likelihood that these events are neutrino signal from observed GRBs.  We use a likelihood function that incorporates the probabilities that an event is a signal neutrino from a GRB or a background atmospheric neutrino or muon.  We calculate these background-like and signal-like probabilities from individually normalized probability distribution functions (PDFs) in time, space, and energy:

\[
\mathcal{S}(\overrightarrow{x_i}) = P^{\mathrm{Time}}_{\mathrm{s}}(t_i) \times P^{\mathrm{Space}}_{\mathrm{s}}(\overrightarrow{r_i}) \times P^{\mathrm{Energy}}_{\mathrm{s}}(E_i)
\]
\[
\mathcal{B}(\overrightarrow{x_i}) = P^{\mathrm{Time}}_{\mathrm{b}}(t_i)  \times P^{\mathrm{Space}}_{\mathrm{b}}(\overrightarrow{r_i}) \times P^{\mathrm{Energy}}_{\mathrm{b}}(E_i)
\]
where \textit{S} and \textit{B} are the probabilities that an event $i$ with properties $\overrightarrow{x}_i$ is signal and background, respectively.

\begin{figure}[h]
 \plotone{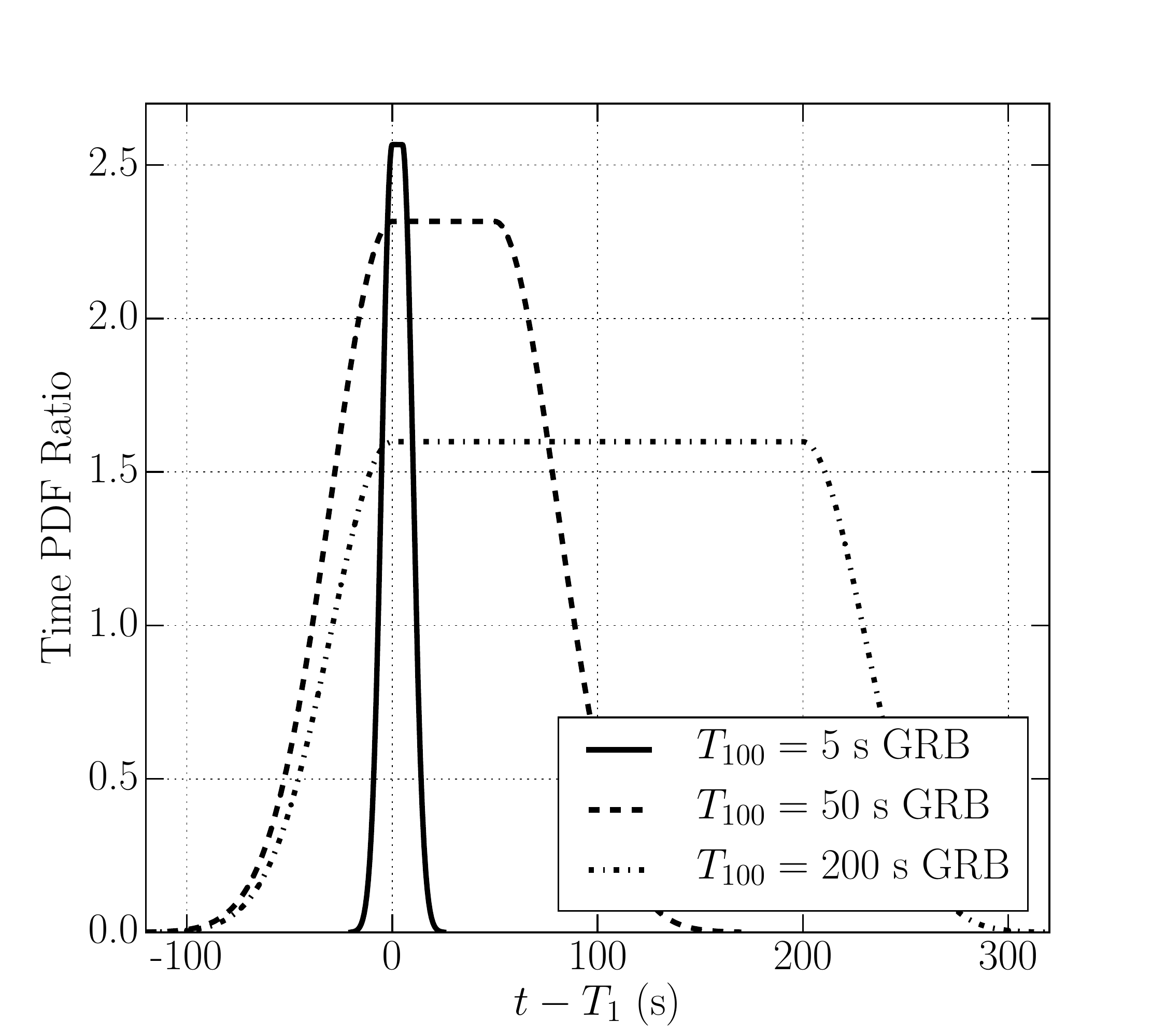}
 \caption{Signal / background time PDF ratios for events during example GRBs with different measured $T_{100}$ values.
 \label{timepdf}}
\end{figure}

The signal time PDF is flat during the gamma-ray emission duration ($T_{100}$ defined in Section \ref{sec:promptgrbnuspec}) and has Gaussian tails before $T_1$ and after $T_2$.  The width of these Gaussian tails $\sigma_{\mathrm{t}}$ equals the duration of measured gamma-ray emission up to 30 s and down to 2 s to account for possible small shifts in the neutrino emission time with respect to that of the photons.  We accept events out to $T_{100}\pm4 \sigma_{\mathrm{t}}$ for each GRB time window.  The background time PDF is flat throughout the total period of acceptance for each GRB.  Examples of the signal / background time PDF ratios for events during GRBs with different measured durations are shown in Figure \ref{timepdf}.

\begin{figure}[h]
  \plottwo{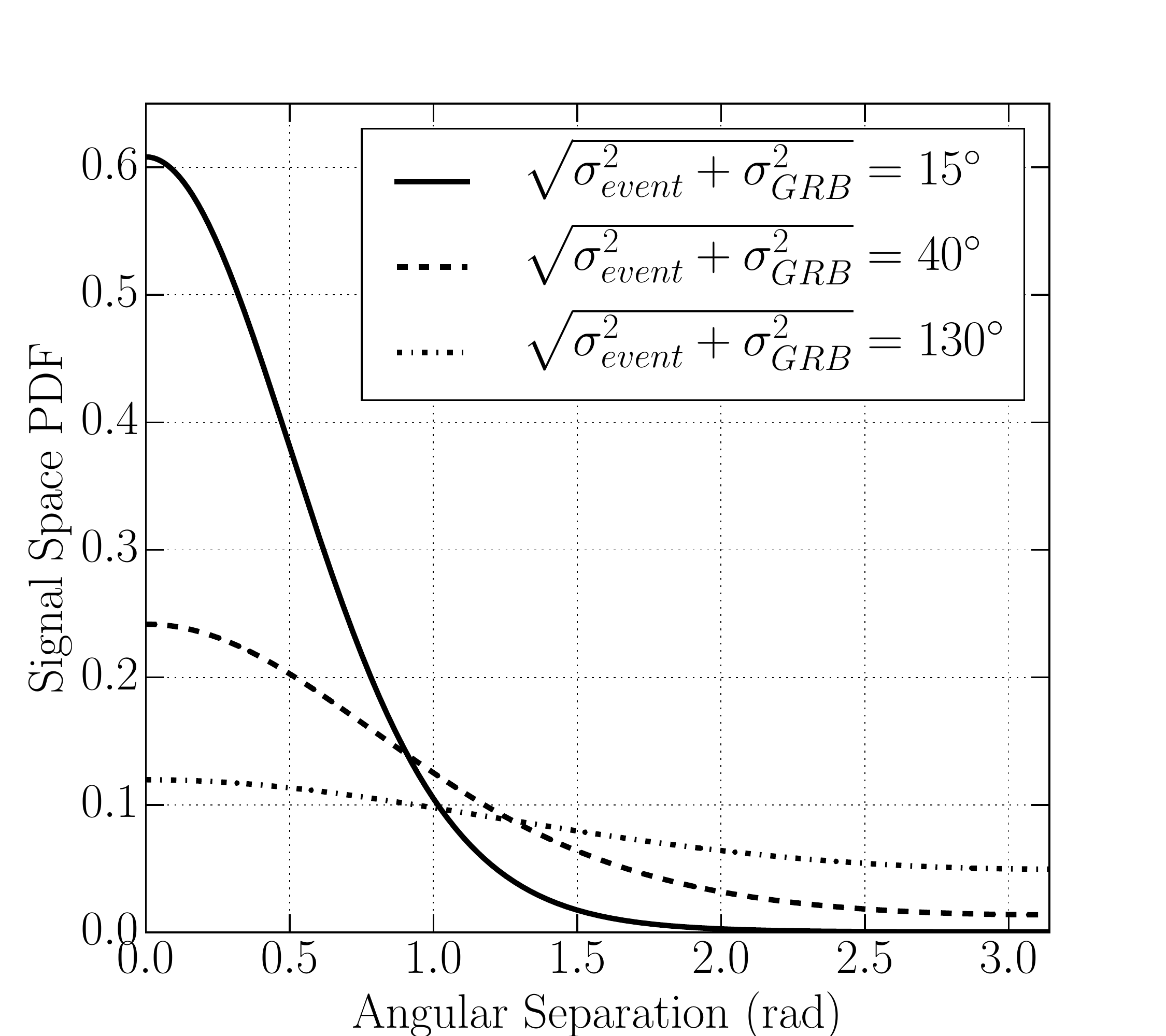}{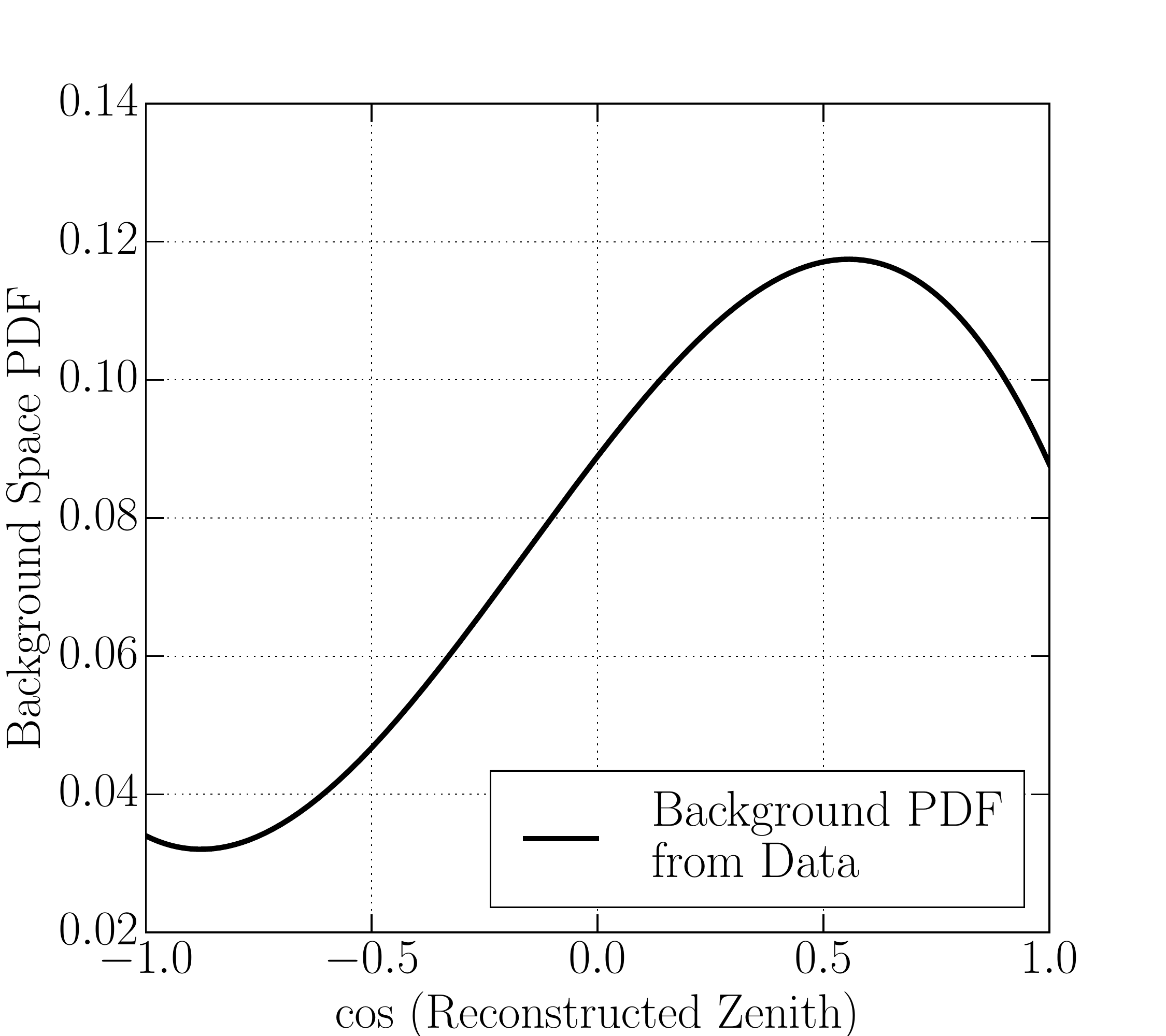}
 \caption{Left: signal space PDFs for three example events and correlated GRBs. Right: background space PDF, calculated from a spline fit to the zenith distribution of data events at final analysis level.  The 79- and 86-string detectors show azimuthal symmetry.
 \label{spacepdf}}
\end{figure}

The signal space PDF is the first-order non-elliptical term in the Kent distribution \citep{kent}:

\begin{equation}
P^{\mathrm{Space}}_{\mathrm{s}}(\overrightarrow{r_i}) = \frac{\kappa}{4\pi\sinh(\kappa)} \e^{\kappa(\hat{r}_i \cdot \hat{r}_{\mathrm{GRB}})}
\label{kent}
\end{equation}
for which the concentration parameter $\kappa=\frac{1}{\sigma^2_{\mathrm{GRB}}+\sigma^2_{i}}$ is the reciprocal of the uncertainty in the GRB's localization and the Cramer-Rao uncertainty in the event's reconstructed direction, $\hat{r}_i$ is the reconstructed direction of the event, and $\hat{r}_{\mathrm{GRB}}$ is the most precise GRB localization available.  We use this spherical analogue of the planar Gaussian distribution because of the large uncertainties in our reconstructed cascade event directions.  The background space PDF is a spline fit to the distribution of reconstructed $\cos(\theta_{\mathrm{zenith}})$ of all final analysis level off-time data events.  Small variance in the background reconstructed azimuth distribution has negligible impact on this analysis, and so our background space PDF only varies with zenith.  Examples of the signal space PDFs for events and correlated GRBs with different directional uncertainties are shown with the background space PDF in Figure \ref{spacepdf}.

\begin{figure}[h]
 \plotone{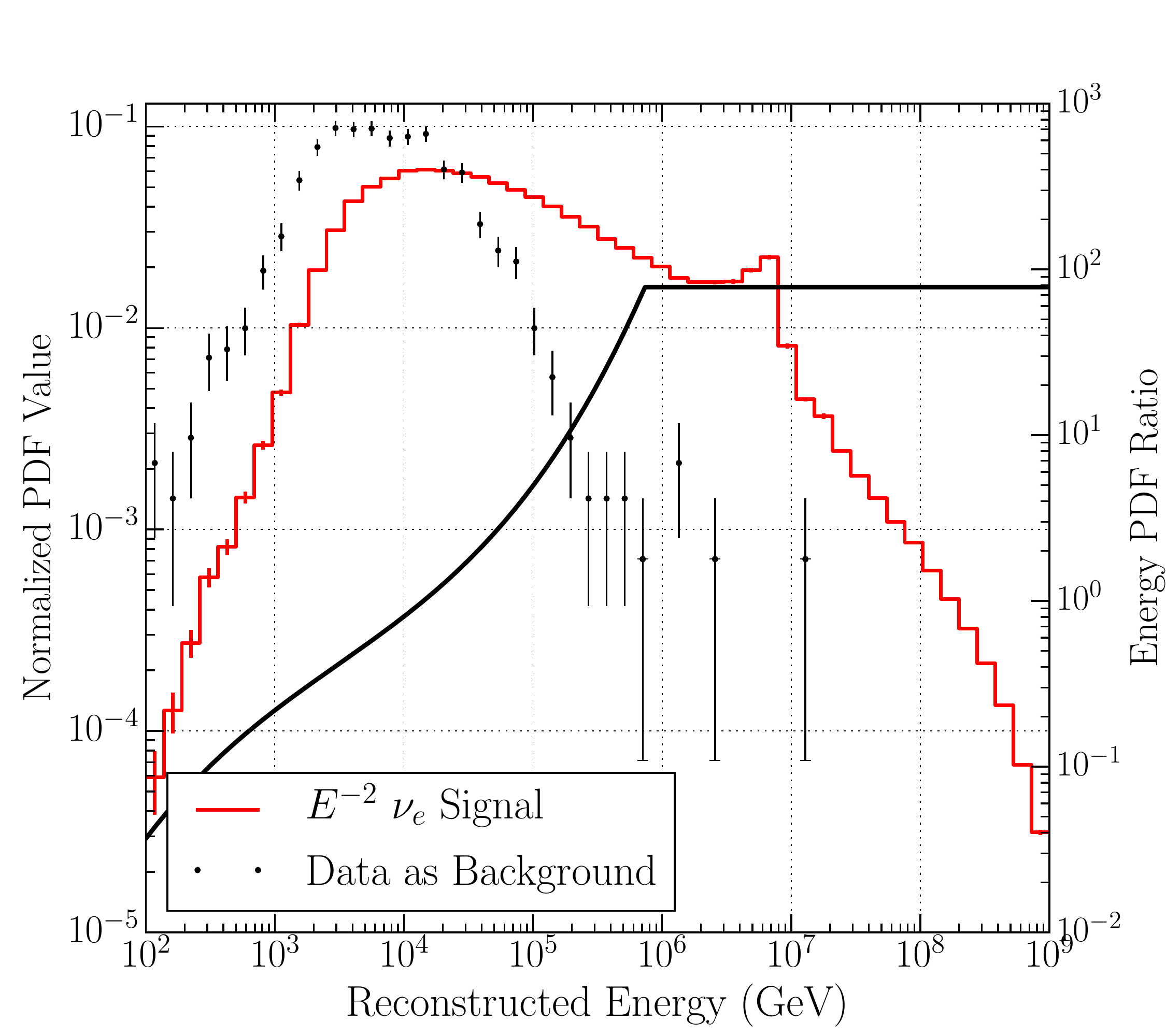}
 \caption{Left vertical axis: Reconstructed energy distributions of data (dot markers) and $\mathrm{E}^{-2}$ $\nu_e$ signal (red line) at final analysis level.  Right vertical axis: Signal / background energy PDF distribution (black line) calculated from a spline fit to the ratio of the two energy distributions.
 \label{energypdf}}
\end{figure}

The signal and background energy PDFs are the reconstructed energy distributions of $E^{-2}$ $\nu_e$ simulation and off-time data, respectively.  We fit a spline to the ratio of these two PDFs.  We find few background events in our final sample that have reconstructed energies above 1~PeV, and so we conservatively assume a constant ratio of signal and background energy PDFs at energies above 1~PeV.  The signal and background energy PDFs and their spline-fit ratio are shown in Figure \ref{energypdf}.

In order to choose our optimal final selection on BDT score and characterize the significance of the result, we construct a test statistic in the form of a maximum likelihood function.  This function incorporates probabilities that observed events are signal and background as well as provides an estimator for the number of observed signal events.  The likelihood function combines the above signal and background PDFs with the Poisson probability $P_N$ of observing $N$ events, given that the expected total number of signal + background events is $n_{\mathrm{s}}+n_{\mathrm{b}}$:

\begin{equation}
\mathcal{L}(\left\{\overrightarrow{x_i}\right\};n_{\mathrm{s}}+n_{\mathrm{b}}) = P_{N} \prod_{i=1}^{N}(p_{\mathrm{s}}\mathcal{S}(\overrightarrow{x_i}) + p_{\mathrm{b}}\mathcal{B}(\overrightarrow{x_i}))
\label{totallh}
\end{equation}
where
\begin{equation}
P_{N} = \frac{(n_{\mathrm{s}}+n_{\mathrm{b}})^N \e^{-(n_{\mathrm{s}}+n_{\mathrm{b}})}}{N!}
\label{poissonlh}
\end{equation}
and $p_{\mathrm{s}}=\frac{n_{\mathrm{s}}}{n_{\mathrm{s}}+n_{\mathrm{b}}}$ and $p_{\mathrm{b}}=\frac{n_{\mathrm{b}}}{n_{\mathrm{s}}+n_{\mathrm{b}}}$ are the probabilities of observing a signal and background event.

We do not know \textit{a priori} the total number of events we will observe.  Therefore, we estimate $n_b$ from the background data rate multiplied by the total search time window $\sum_{i}^{N_{\mathrm{GRBs}}} (T_{100, i} + 8\sigma_{\mathrm{t},i})$ in which we accept events; and this expectation is denoted by $\langle n_{\mathrm{b}} \rangle$.  We estimate $n_{\mathrm{s}}$ from the value $\hat{n}_{\mathrm{s}}$ that maximizes $\mathcal{L}$.  Additionally, we simplify the likelihood function by dividing by its background-only hypothesis and taking the logarithm of this ratio without loss of generality.  Finally, our test statistic $T$ is the maximized value of this ratio at $\hat{n}_{\mathrm{s}}$:

\begin{equation} \label{teststat}
T = -\hat{n}_{\mathrm{s}} + \sum_{i=0}^N\ln{\left[\frac{\hat{n}_{\mathrm{s}} \mathcal{S}(\overrightarrow{x_i})}{\langle n_{\mathrm{b}} \rangle \mathcal{B}(\overrightarrow{x_i})} + 1\right]}
\end{equation}
We extend to multiple detector configurations (79~strings, first year of 86 strings, second year of 86 strings, etc.) and search channels (cascades, tracks) by adding maximized test statistics for each configuration and channel:

\begin{equation}
T = \sum_{\mathrm{c}}\left\{-(\hat{n}_{\mathrm{s}})_{\mathrm{c}} + \sum_{i=0}^{N_{\mathrm{c}}}\ln{\left[\frac{(\hat{n}_{\mathrm{s}})_{\mathrm{c}} \mathcal{S}_{\mathrm{c}}(\overrightarrow{x_i})}{\langle n_{\mathrm{b}} \rangle_{\mathrm{c}} \mathcal{B}_{\mathrm{c}}(\overrightarrow{x_i})} + 1\right]}\right\}
\label{teststatcomb}
\end{equation}
where c represents each combination of search channel and detector configuration.

\begin{figure}[h]
 \plotone{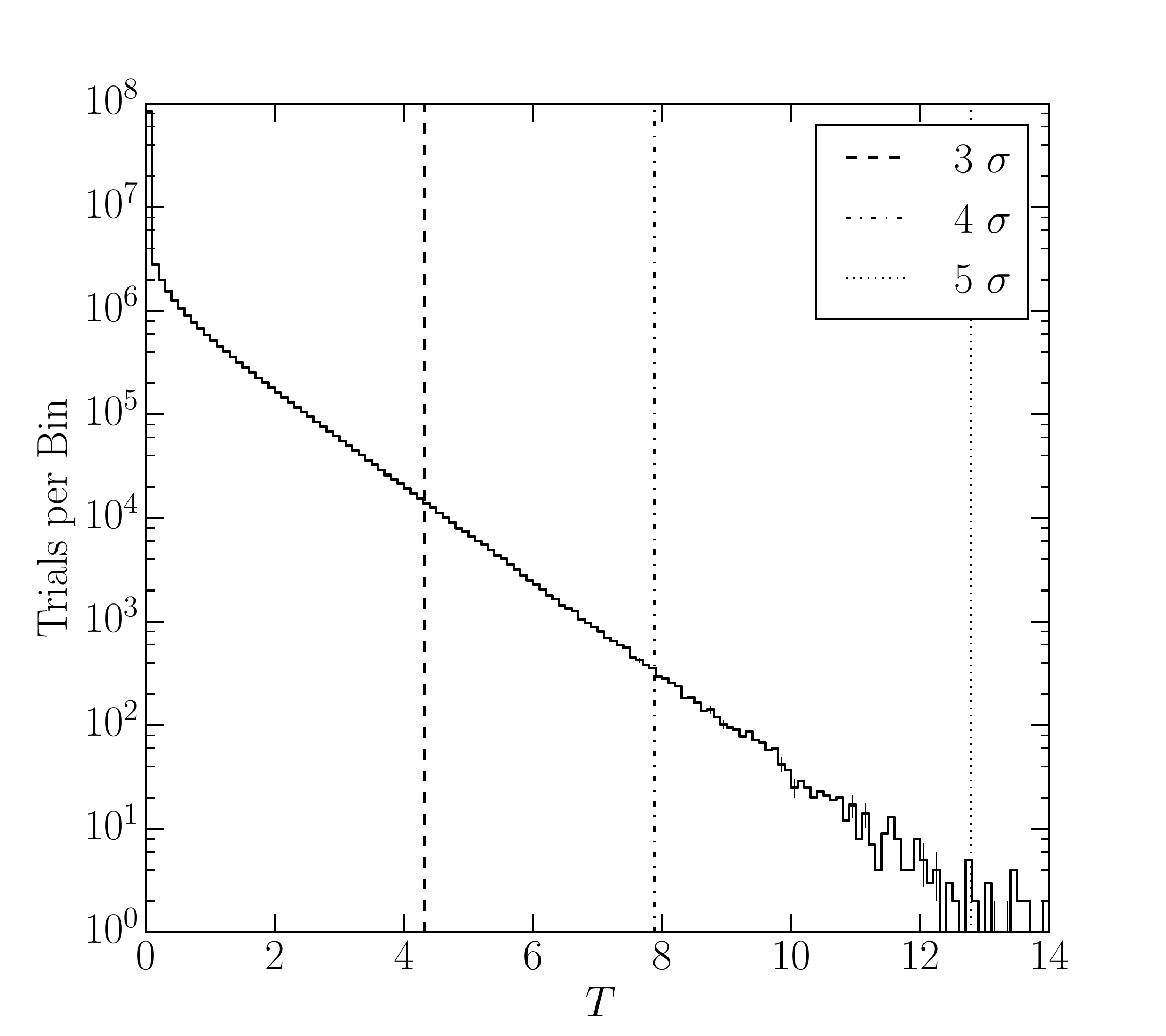}
 \caption{Test statistic distribution for $10^8$ randomized background-only pseudo-searches at final analysis level.  The vertical lines represent test statistic values for a $3\sigma$, $4\sigma$, and $5\sigma$ discovery.
 \label{tsd}}
\end{figure}

To set discovery significance thresholds, we first perform $10^8$  pseudo-search trials using only background data for a range of BDT score $> s$ cuts.  Each background sample has its own $\langle n_{\mathrm{b}} \rangle_{\mathrm{cut}}$, with lower values for cuts on higher BDT scores.  In each background-only trial, for each GRB, we choose a pseudo-random number of observed events within our $T_{100} \pm 4\sigma_{\mathrm{t}}$ search time window about the gamma-ray emission from a Poisson distribution with expectation determined by the background data rate and time window.  If the number of events is 0, then $T$ receives no contribution from the search window about that GRB.  If the number of events is greater than 0, then we construct each event using the following steps: (1) choose a random time PDF value; (2) choose a random azimuth within 0 to $2\pi$; (3) choose a reconstructed energy by sampling from the background distribution; (4) choose a reconstructed zenith by sampling from the background distribution of events with similar energy; (5) choose an estimated error in reconstructed direction by sampling from the background distribution of events with similar energy and zenith.  Finally, with the signal and background PDF values for every event, we calculate the test statistic for each trial.  We obtain a distribution like that shown in Figure \ref{tsd} for each possible selection on BDT score.

\begin{figure}[h]
  \plottwo{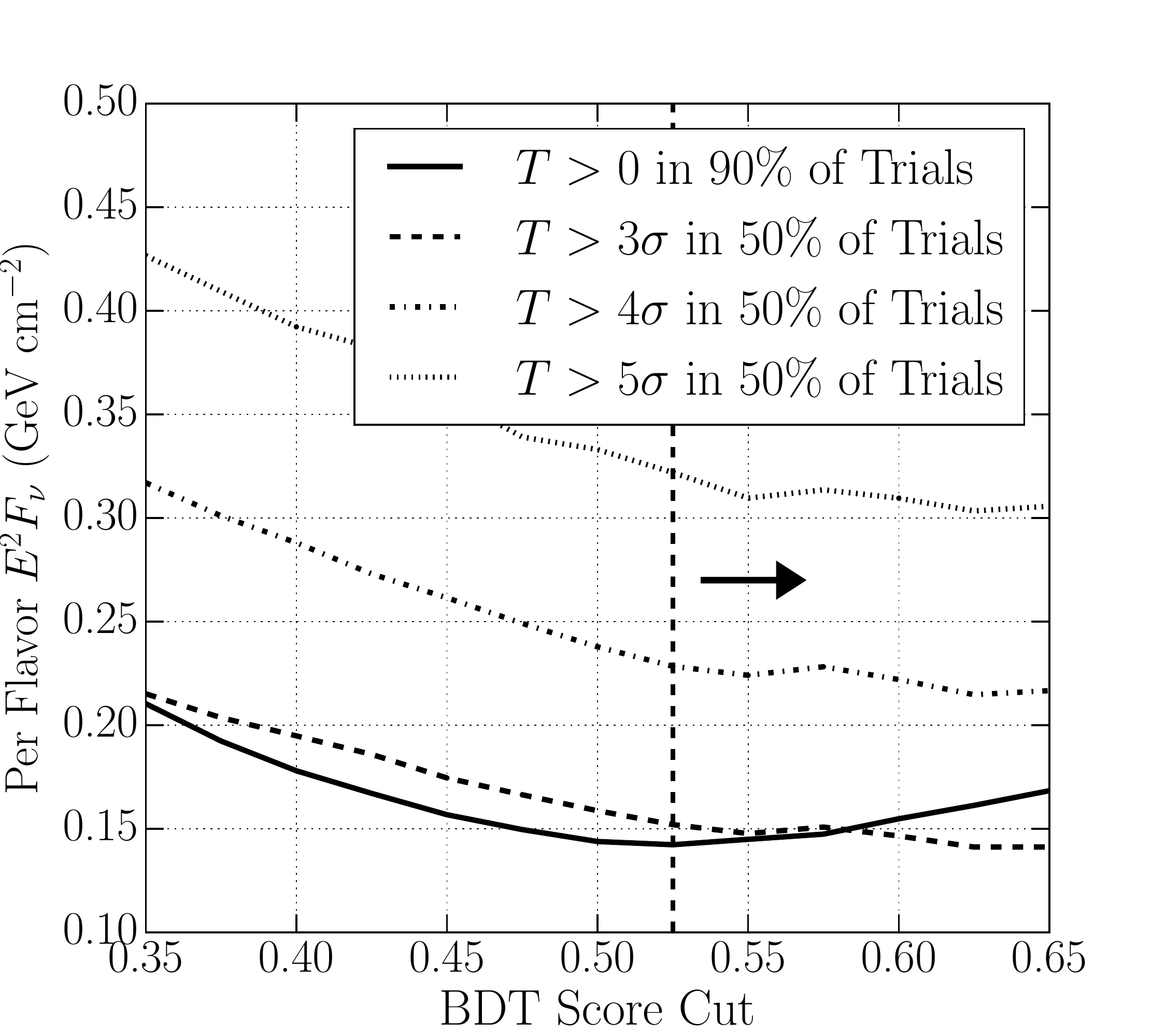}{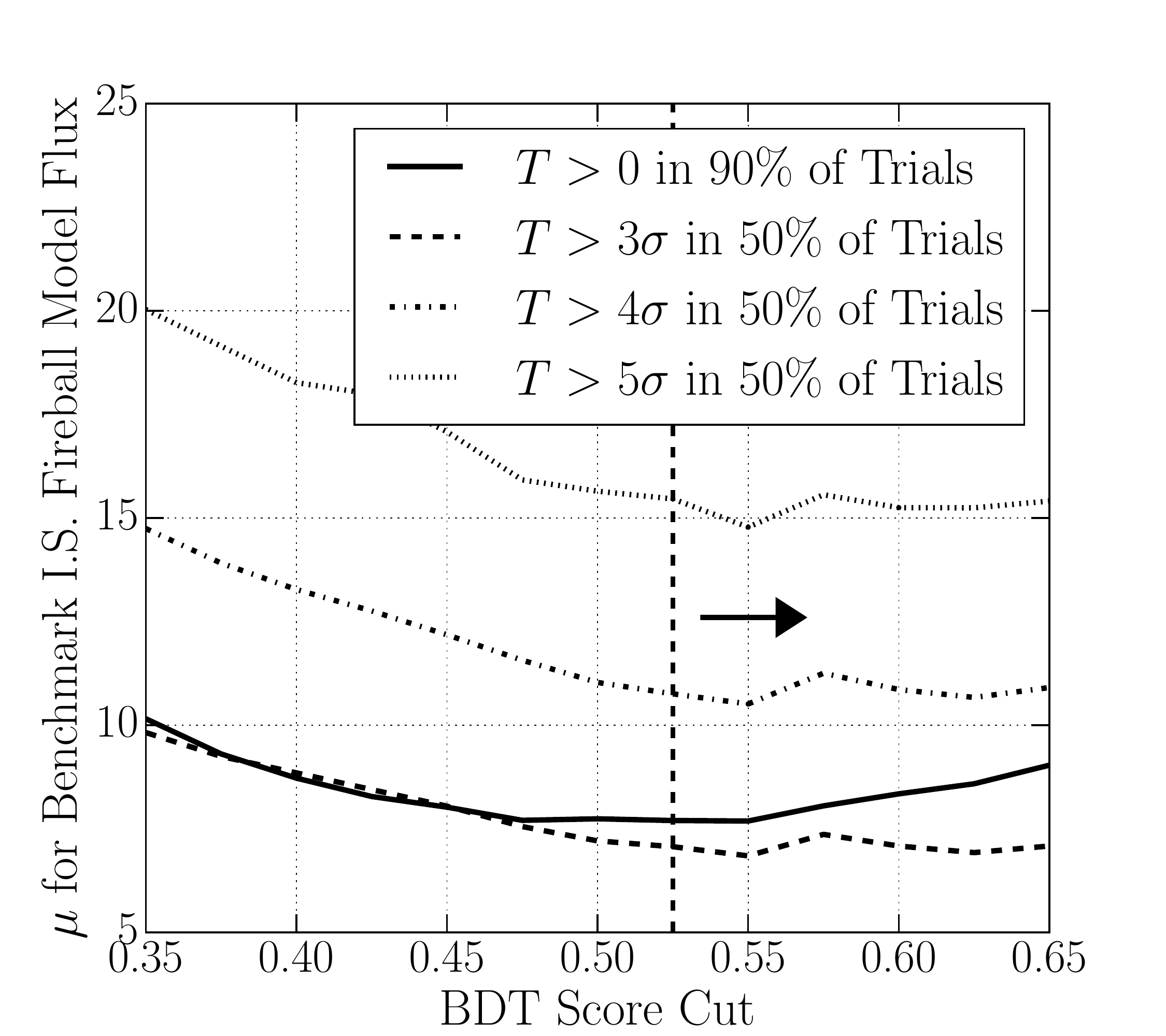}
 \caption{Limit setting and discovery potentials per cut on BDT score for a single year's search.  Horizontal axis corresponds to a BDT score $> s$ cut.  The final event selection was optimized for limit setting while suffering little loss in discovery potential.  The vertical dashed line represents our final event selection of BDT score $> 0.525$.  Left: the vertical axis corresponds to the $E^{-2}$ spectrum signal weight needed in order to reach the given threshold.  Right: the vertical axis corresponds to the multiplying factor $\mu$ on the benchmark internal shock fireball model neutrino flux shown in Figure \ref{grbspectra} in order to reach the given threshold.
 \label{llhoptimization}}
\end{figure}

We choose our optimal final selection on BDT score by injecting simulated neutrino signal along with background data and performing $10^4$ pseudo-search trials for a range of BDT score $> s$ cuts.  The background events are selected for each GRB the same as above for the background-only trials.  The simulated signal events within an $11^{\circ}$ circle about each GRB contribute to the likelihood with a probability proportional to their simulated weight.  Signal is increasingly weighted until a defined discovery or limit-setting threshold is reached.  These discovery and limit-setting potentials per cut on BDT score are shown in Figure~\ref{llhoptimization} for a general $E^{-2}$ spectrum and the Figure~\ref{grbspectra} benchmark internal shock fireball spectrum.  The final selection, BDT score $> 0.525$, was optimized to set the best possible upper limit while suffering little loss in discovery potential.  This optimization was performed for each search year's event selection.

\section{Results}
\label{sec:results}

In three years of data, five cascade neutrino candidate events are correlated with five GRBs.  These coincidences yield a combined \textit{P}-value of 0.32, derived from the total test statistic value with respect to the background-only distribution.  The Northern Hemisphere track searches in four years of data resulted in a single neutrino candidate event correlated with prompt GRB emission and a combined \textit{P}-value of 0.46 \citep{grb4year}.  Considering the atmospheric neutrino purity of each search, discussed in Section \ref{sec:eventselect}, the track event is almost certainly a $\nu_{\mu}$ while the cascade events could be high energy atmospheric muons.

Table \ref{tab:results} shows the time, space, and energy data for these events and GRBs.  Events 1, 2, and 4 occurred on the edge or corner of the detector.  Events 3 and 5 occurred within the instrumented volume.  The large directional uncertainties of the second and fourth events are due to the location of their interactions at the corner and edge of the instrumented volume, with relatively few DOMs able to record the Cherenkov light.  The reconstructed energy of the muon neutrino track coincidence discussed in \cite{grb4year} is an approximate lower bound on the neutrino energy because the interaction could have taken place well outside of IceCube and consequently, the muon could have lost significant energy before reaching the detector.

\begin{table*}
\begin{center}
	\begin{tabular}{ccccc} \toprule
	   & Time & Angular Uncertainty & Angular Separation & Fluence/Energy \\ \midrule
	 GRB 101213A & $T_{100} = 202$ s & $0.0005^{\circ}$ & & $7.4 \times 10^{-6}$ erg cm$^{-2}$ \\
	 Event 1 & $T_1 + 109$ s & $32.0^{\circ}$ & $23^{\circ}$ & 11 TeV \\ \midrule
	 GRB 110101B & $T_{100} = 235$ s & $16.5^{\circ}$ & & $6.6 \times 10^{-6}$ erg cm$^{-2}$ \\
	 Event 2 & $T_1 + 141$ s & $118^{\circ}$ & $112^{\circ}$ & 34 TeV \\ \midrule
	 GRB 110521B & $T_{100} = 6.14$ s & $1.31^{\circ}$ & & $3.6 \times 10^{-6}$ erg cm$^{-2}$ \\
	 Event 3 & $T_1 + 0.26$ s & $16.5^{\circ}$ & $35^{\circ}$ & 3.4 TeV \\ \midrule
	 GRB 111212A & $T_{100} = 68.5$ s & $0.0004^{\circ}$ & & $1.4 \times 10^{-6}$ erg cm$^{-2}$ \\
	 Event 4 & $T_1 + 11.7$ s & $44.8^{\circ}$ & $120^{\circ}$ & 31 TeV \\ \midrule
	 GRB 120114A & $T_{100} = 43.3$ s & $0.04^{\circ}$ & & $2.4 \times 10^{-6}$ erg cm$^{-2}$ \\
	 Event 5 & $T_1 + 57.2$ s & $7.9^{\circ}$ & $18^{\circ}$ & 3.8 TeV \\ \midrule
	 GRB 100718A$^a$ & $T_{100} = 39$ s & $10.2^{\circ}$ & & $2.5 \times 10^{-6}$ erg cm$^{-2}$ \\
	 $\nu_{\mu}$ Track Event$^a$ & $T_1 + 15$ s & $1.3^{\circ}$ & $16^{\circ}$ & $\gtrsim$ 10 TeV \\
	 \bottomrule
	\end{tabular}
     \caption{GRB and Event Properties for the Three-year Cascade and Four-year Track Search Coincidences.\newline$^a$ Corresponds to the $\nu_{\mu}$ track search coincidence discussed in \cite{grb4year}.
     \label{tab:results}}
\end{center}
\end{table*}

The benchmark internal shock, photospheric, and ICMART fireball model spectra plotted in Figure \ref{grbspectra} yield expectations of 1.5, 2.5, and 0.07 neutrinos, respectively, for this three-year all-flavor cascade analysis.  As discussed at the end of Section 5, this search achieves nearly the same sensitivity as the Northern Hemisphere track search.  Thus, the combined four years of track searches and three years of shower searches yield total benchmark model expectations of 3.3, 5.4, and 0.1 neutrinos.  The all-flavor shower search has an average $\langle n_{\mathrm{b}} \rangle$ of 11 events per year.  This expectation is concentrated at lower energies than the expected signal and is weighted by the time, space, and energy PDFs accordingly in our unbinned likelihood.  The expected background during just the $T_{100}$ of each GRB is 3.5 events per year.  An observation of three 1~PeV neutrinos correlated with three GRBs, with the same temporal and spatial properties as our observed Event 1 and its respective GRB in Table \ref{tab:results}, would yield a $T$ value over 12 and a $5\sigma$ discovery based on the background-only distribution in Figure \ref{tsd}.

Considering the low significance of the events correlated with GRBs, we set 90\% confidence level (CL) Neyman upper limits \citep{neyman1937} on models normalized to the observed flux of UHECRs as well as models normalized to the observed gamma-ray fluence of each GRB.  We calculate these limits by combining the three-year cascade search results and four-year Northern Hemisphere track search results using the multiple configuration and channel test statistic given in Equation \ref{teststatcomb}.  Simulated neutrinos are weighted to a certain spectrum and normalization and injected over background in the pseudo-searches.  The exclusion CL is the fraction of pseudo-search trials that yield $T \geq T_{\mathrm{observed}}$.

\begin{figure}[h]
 \plotone{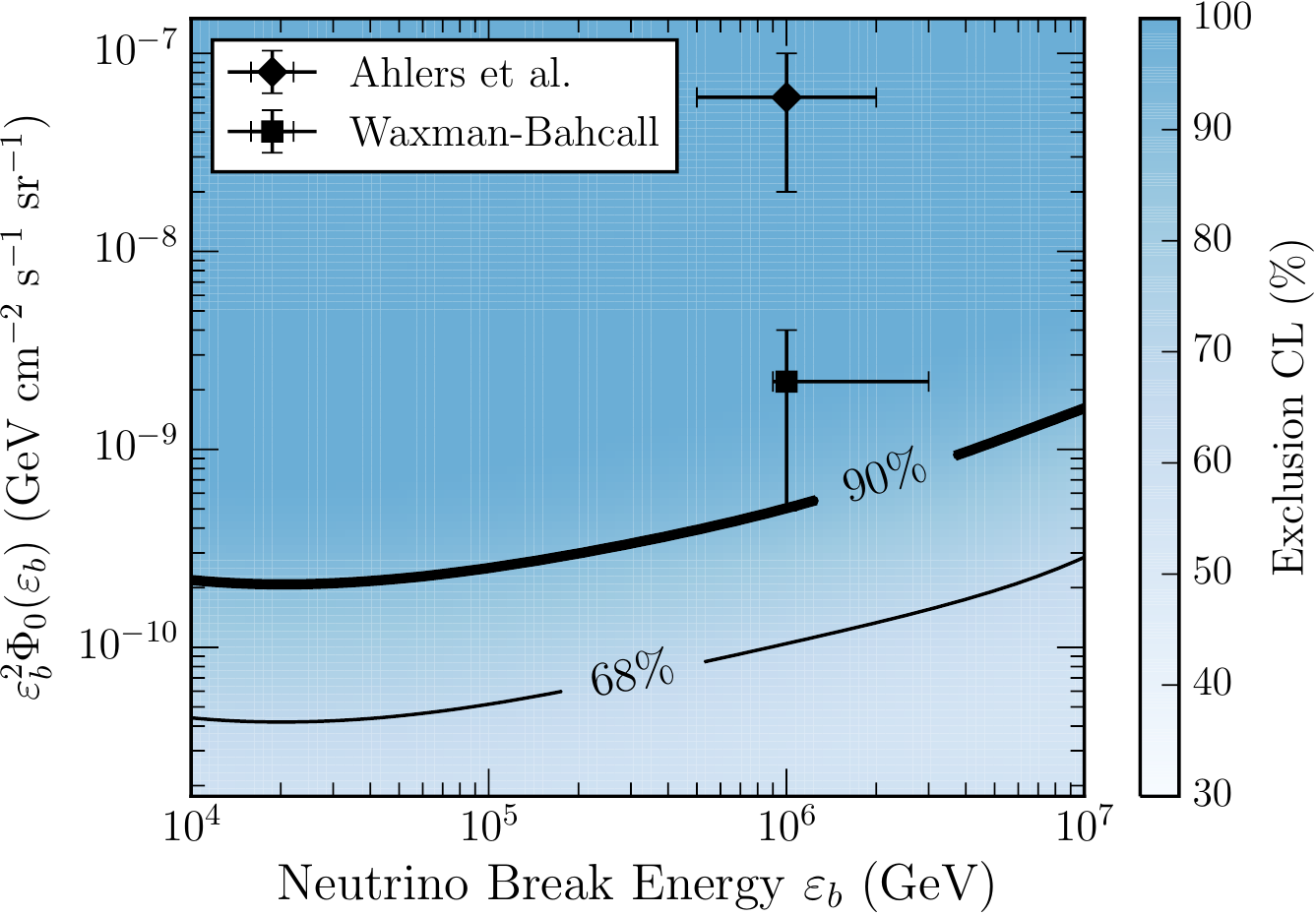}
 \caption{Exclusion contours, calculated from the combined three-year all-sky $\nu_{e}$ $\nu_{\tau}$ $\nu_{\mu}$ shower-like event search and four-year Northern Hemisphere $\nu_{\mu}$ track-like event search results, of a per-flavor double broken power law GRB neutrino flux of a given flux normalization $\phi_0$ at first break energy $\epsilon_b$.
 \label{topdownlimits}}
\end{figure}

Figure \ref{topdownlimits} shows exclusion contours for double broken power law spectra of the form

\begin{equation}
\begin{aligned}
\Phi_{\nu}(E) = \phi_0 \times \left\{
	\begin{array}{ll}
	E^{-1} \epsilon_b^{-1}, & E < \epsilon_b \\
	E^{-2}, & \epsilon_b \leq E < 10\epsilon_b \\
	E^{-4} (10\epsilon_b)^2, & 10\epsilon_b \leq E
	\end{array}
\right.
\end{aligned}
\label{doublebrokenplaw}
\end{equation}
at different first break energies $\epsilon_b$ and normalizations $\epsilon_b^2 \phi_0$.  Our combined limits largely rule out cosmic ray escape via neutron production \citep{probation2011}.  Mechanisms allowing for cosmic ray escape via protons \citep{wb1997} are disfavored as well.  The \cite{wb1997} model has been updated to account for more recent measurements of the UHECR flux \citep{katz2009energy} and typical gamma break energy \citep{goldstein2012}.

\begin{figure}[h]
  \plotthree{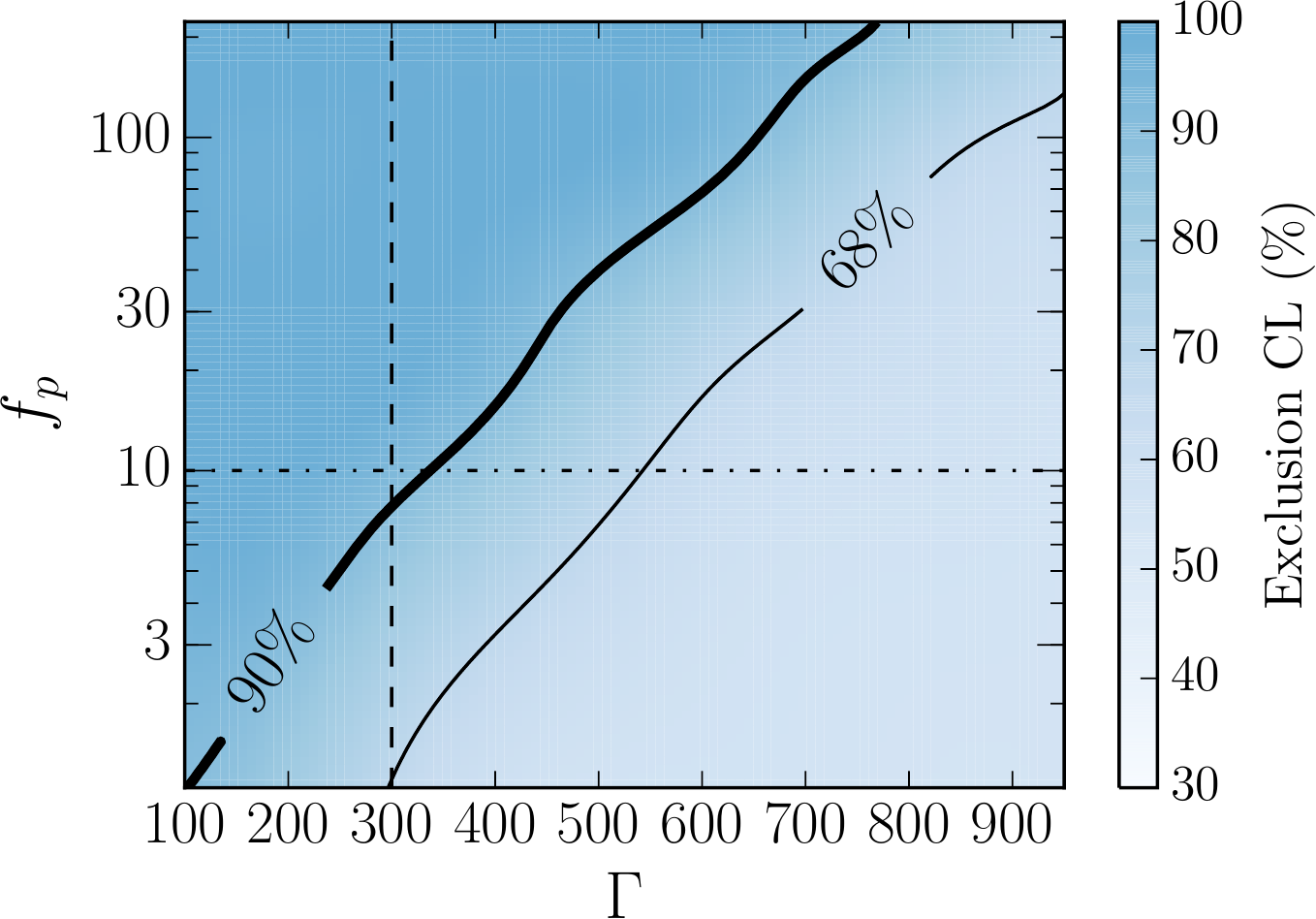}{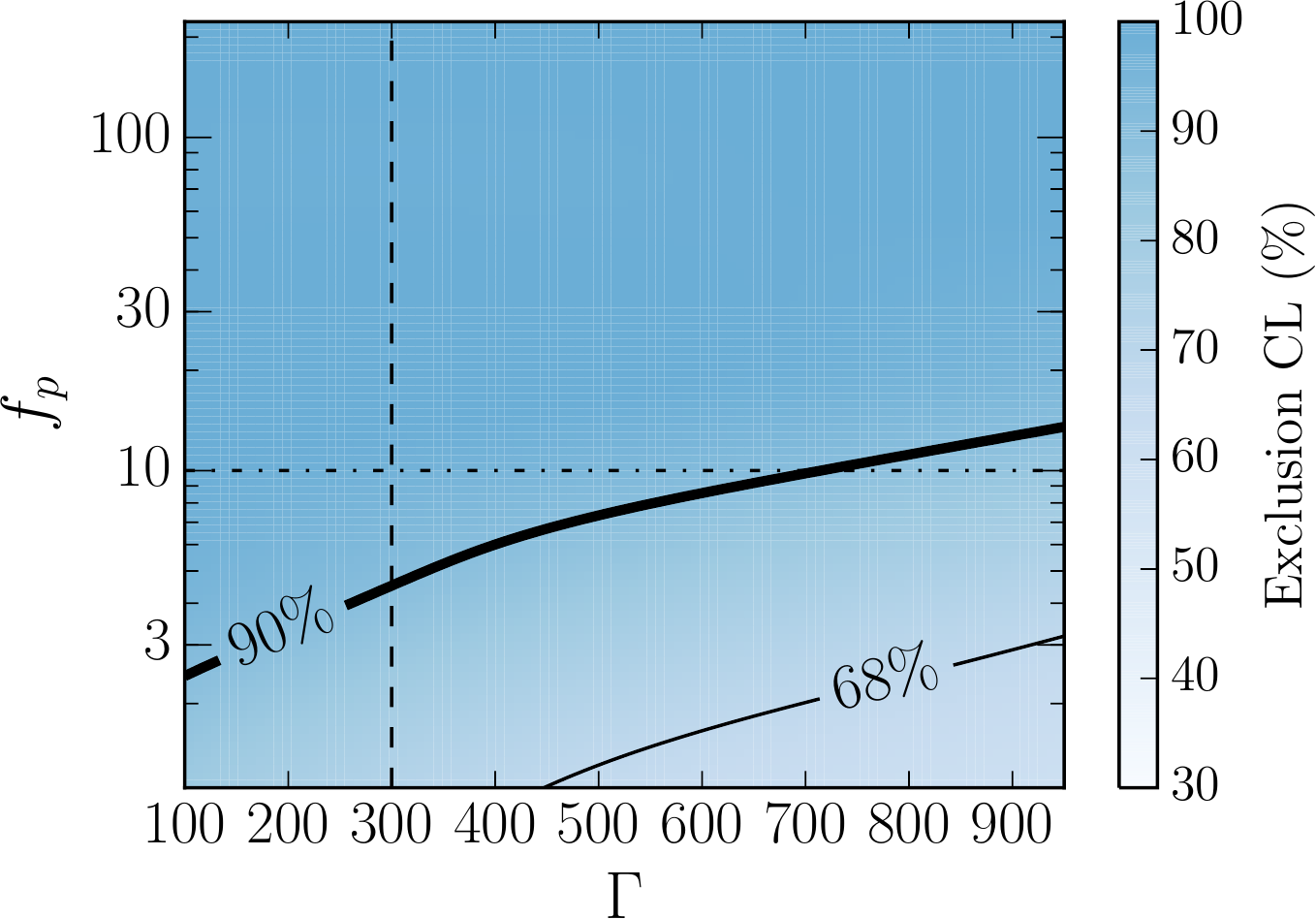}{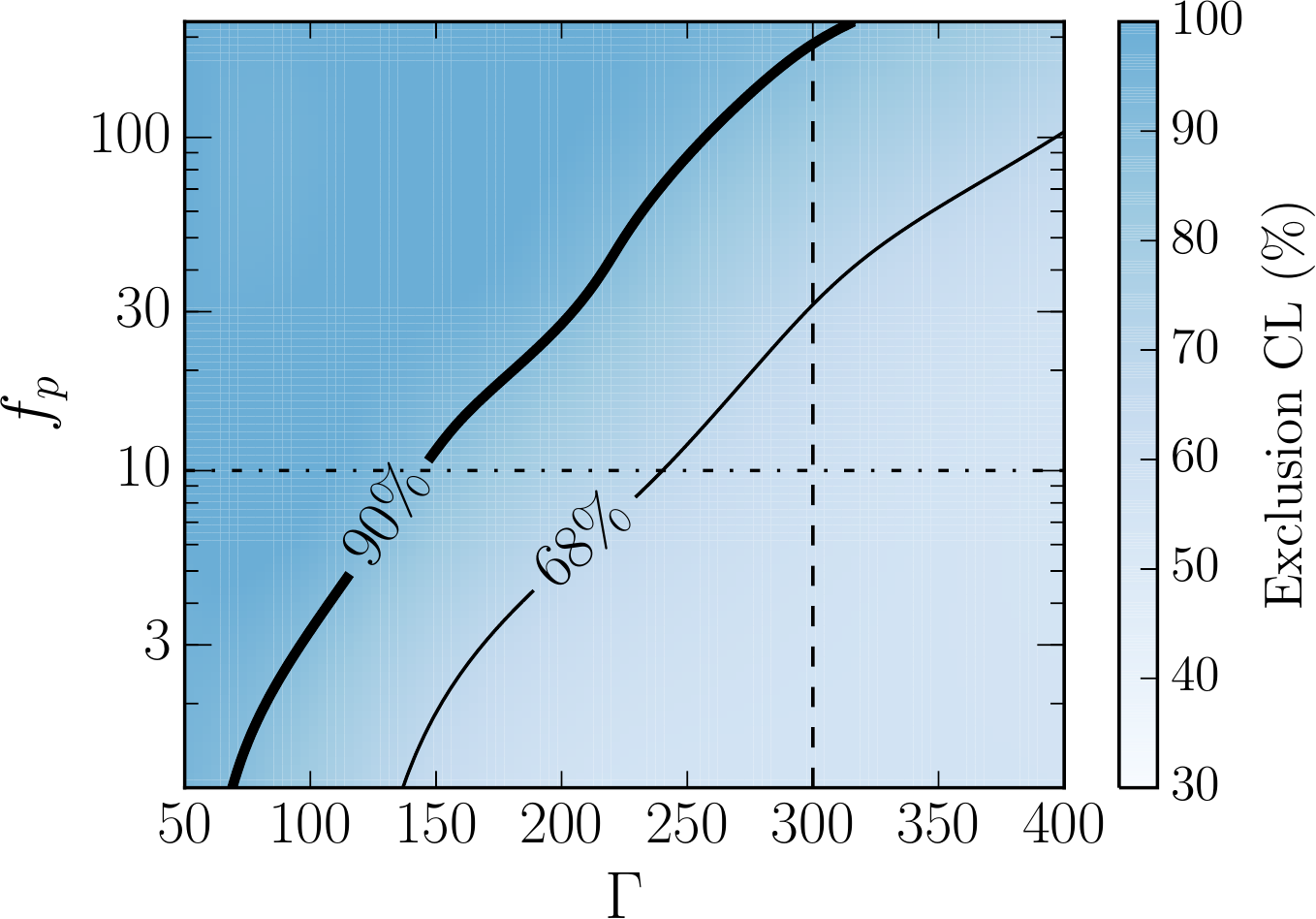}
 \caption{Exclusion contours, calculated from the combined three-year all-sky $\nu_{e}$ $\nu_{\tau}$ $\nu_{\mu}$ shower-like event search and four-year Northern Hemisphere $\nu_{\mu}$ track-like event search results, in $f_p$ and $\Gamma$ GRB parameter space, for three different models of fireball neutrino production. These models differ in the radius at which $p\gamma$ interactions occur.  The vertical and horizontal dashed lines indicate the benchmark parameters used for Figure \ref{grbspectra} and the right panel of Figure \ref{llhoptimization}.  Left: internal shock.  Middle: photospheric.  Right: ICMART.
 \label{bottomuplimits}}
\end{figure}

Figure \ref{bottomuplimits} shows exclusion contours in the baryonic loading - bulk Lorentz factor parameter space for the internal shock, photospheric, and ICMART per-burst gamma-ray-normalized fireball models.  The benchmark model spectra from Figure \ref{grbspectra} are indicated by the intersection of the vertical and horizontal dashed lines.  Systematic uncertainties in the ice model, DOM photon detection efficiency, and lepton propagation in the earth and ice are added in quadrature to a total effect of $\sim11\%$ for each model limit.  The addition of the all-sky cascade search over three years of data and GRBs improves the Northern Hemisphere track search limits by about 50\%.

\section{Conclusions and Outlook}
\label{sec:outlook}

We have performed a search for neutrinos of all flavors emitted by 807 GRBs during three years of IceCube data.  This search exhibits similar sensitivity to that of previously published searches for muon neutrinos emitted by 506 Northern Hemisphere GRBs during four years of IceCube data.  Over both search channels, we found six events that are correlated with GRBs but also consistent with background.  Our limits placed on neutrino emission models normalized to the observed UHECR flux are the strongest constraints thus far on the hypothesis that GRBs are the dominant sources of this flux.  Additionally, our limits placed on the latest neutrino emission models normalized to the observed gamma-ray fluence from each GRB constrain parts of the parameter space relevant for the production of UHECR protons.  Models that are still allowed require increasingly lower neutrino production efficiencies through large bulk Lorentz boost factors, low baryonic loading, or large dissipation radii.

As shown in \cite{baerwald2014}, constraints on parameters involved in fireball neutrino production via internal shock collisions can be connected to the requirement that GRBs are the sources of the observed UHECRs in a self-consistent way, assuming a pure proton composition.  While cosmic ray escape from GRB fireballs through neutron production is strongly constrained by our limits, \cite{baerwald2014} show that the unexcluded parameter space in Figure \ref{bottomuplimits} allows for efficient diffusive proton escape assuming a galactic-to-extragalactic source transition at the ankle of the cosmic ray spectrum.  Although the allowed parameter space of this model is plausible, the average burst likely exhibits $\Gamma$ and $f_p$ values that are largely excluded for neutrino production.  Nevertheless, we have only considered single-zone neutrino and cosmic ray emission models of GRBs in this work, and models of multiple emission regions predict a neutrino flux below our current sensitivity~\citep{multipleis2015, globus2015uhecr}.

Our acceptance of possible prompt neutrino signal can increase further through the addition of searches for $\nu_{\mu}$-induced tracks correlated with Southern Hemisphere GRBs that occurred during the 79-string and completed detector configurations.  Moreoever, the next-generation \textit{IceCube-Gen2} detector will significantly improve the sensitivity to transient sources \citep{icecubegen2, AhlersHalzenicecubegen2}.  The continued pursuit of all neutrino flavors from observed GRBs over the entire sky will either reveal a flux that is still lower than our current sensitivity or increasingly disfavor these phenomena as sources of the highest energy cosmic rays.

\acknowledgments

We acknowledge the support from the following agencies:
U.S. National Science Foundation-Office of Polar Programs,
U.S. National Science Foundation-Physics Division,
University of Wisconsin Alumni Research Foundation,
the Grid Laboratory Of Wisconsin (GLOW) grid infrastructure at the University of Wisconsin - Madison, the Open Science Grid (OSG) grid infrastructure;
U.S. Department of Energy, and National Energy Research Scientific Computing Center,
the Louisiana Optical Network Initiative (LONI) grid computing resources;
Natural Sciences and Engineering Research Council of Canada,
WestGrid and Compute/Calcul Canada;
Swedish Research Council,
Swedish Polar Research Secretariat,
Swedish National Infrastructure for Computing (SNIC),
and Knut and Alice Wallenberg Foundation, Sweden;
German Ministry for Education and Research (BMBF),
Deutsche Forschungsgemeinschaft (DFG),
Helmholtz Alliance for Astroparticle Physics (HAP),
Research Department of Plasmas with Complex Interactions (Bochum), Germany;
Fund for Scientific Research (FNRS-FWO),
FWO Odysseus programme,
Flanders Institute to encourage scientific and technological research in industry (IWT),
Belgian Federal Science Policy Office (Belspo);
University of Oxford, United Kingdom;
Marsden Fund, New Zealand;
Australian Research Council;
Japan Society for Promotion of Science (JSPS);
the Swiss National Science Foundation (SNSF), Switzerland;
National Research Foundation of Korea (NRF);
Danish National Research Foundation, Denmark (DNRF)

\bibliographystyle{apj}
\bibliography{bibdatabase}



\end{document}